\documentclass[12pt]{iopart}
\usepackage{iopams}
\usepackage{xcolor}
\usepackage{colortbl}
\usepackage{pstricks,multido}
\usepackage{wrapfig}
\usepackage{graphicx,epsfig}
\usepackage{paralist}
\usepackage[margin=2.5cm,top=2cm,bottom=2cm,centering]{geometry}

\definecolor{lightyellow}{rgb}{.95,.96,.5}
\definecolor{myblue}{rgb}{.39,.39,0.92}
\definecolor{midblue}{rgb}{.7,.7,1}
\definecolor{lightblue}{rgb}{.8,.8,1}
\definecolor{lightblue2}{rgb}{.9,.9,1}
\definecolor{macblue}{rgb}{.40,.55,0.84}
\definecolor{mygray}{rgb}{.75,.75,.75}
\definecolor{myred}{rgb}{1,0.61,0.61}
\definecolor{lightred}{rgb}{0.82,0.23,0.13}
\definecolor{lightred2}{rgb}{1.,0.8,0.8}
\definecolor{mygreen}{rgb}{0.40,0.75,0.16}
\definecolor{mybrown}{rgb}{0.69,0.49,0.30}
\definecolor{lightgreen}{rgb}{0.66,1.0,0.67}

\def\twotwo{\setlength{\unitlength}{0.3mm}
\begin{picture}(25,8)(-5,-4)
\put(0,0){\circle{10}}
\put(0,0){\makebox(0,0)[c]{\scriptsize 2}}
\put(5,0){\line(1,0){5}}
\put(15,0){\circle{10}}
\put(15,0){\makebox(0,0)[c]{\scriptsize 2}}
\put(-7,0){\line(1,0){2}}
\put(20,0){\line(1,0){2}}
\put(0,5){\line(0,1){2}}
\put(15,5){\line(0,1){2}}
\put(0,-5){\line(0,-1){2}}
\put(15,-5){\line(0,-1){2}}
\end{picture}}

\def\abcd{\setlength{\unitlength}{0.3mm}
\begin{picture}(57,8)(-6,-4)
\put(0,0){\circle{10}}
\put(0,0){\makebox(0,0)[c]{\scriptsize $a$}}
\put(5,0){\line(1,0){5}}
\put(15,0){\circle{10}}
\put(14.7,1){\makebox(0,0)[c]{\scriptsize $b$}}
\put(20,0){\line(1,0){5}}
\put(30,0){\circle{10}}
\put(29.7,0){\makebox(0,0)[c]{\scriptsize $c$}}
\put(35,0){\line(1,0){5}}
\put(45,0){\circle{10}}
\put(44.5,1){\makebox(0,0)[c]{\scriptsize $d$}}
\put(-7,0){\line(1,0){2}}
\put(50,0){\line(1,0){2}}
\put(0,5){\line(0,1){2}}
\put(15,5){\line(0,1){2}}
\put(30,5){\line(0,1){2}}
\put(45,5){\line(0,1){2}}
\put(0,-5){\line(0,-1){2}}
\put(15,-5){\line(0,-1){2}}
\put(30,-5){\line(0,-1){2}}
\put(45,-5){\line(0,-1){2}}
\end{picture}}

\def\2342{\setlength{\unitlength}{0.3mm}
\begin{picture}(57,8)(-6,-4)
\put(0,0){\circle{10}}
\put(0,0){\makebox(0,0)[c]{\scriptsize 2}}
\put(5,0){\line(1,0){5}}
\put(15,0){\circle{10}}
\put(15,0){\makebox(0,0)[c]{\scriptsize 3}}
\put(20,0){\line(1,0){5}}
\put(30,0){\circle{10}}
\put(30,0){\makebox(0,0)[c]{\scriptsize 4}}
\put(35,0){\line(1,0){5}}
\put(45,0){\circle{10}}
\put(45,0){\makebox(0,0)[c]{\scriptsize 2}}
\put(-7,0){\line(1,0){2}}
\put(50,0){\line(1,0){2}}
\put(0,5){\line(0,1){2}}
\put(15,5){\line(0,1){2}}
\put(30,5){\line(0,1){2}}
\put(45,5){\line(0,1){2}}
\put(0,-5){\line(0,-1){2}}
\put(15,-5){\line(0,-1){2}}
\put(30,-5){\line(0,-1){2}}
\put(45,-5){\line(0,-1){2}}
\end{picture}}

\def\conf#1{
\ifnum#1=0 
{\begin{picture}(5,8)(-3,-3)
\put(0,0){\circle{10}}
\put(0,0){\makebox(0,0)[c]{\scriptsize 1}}
\put(-7,0){\line(1,0){2}}
\put(5,0){\line(1,0){2}}
\put(0,5){\line(0,1){2}}
\put(0,-5){\line(0,-1){2}}
\end{picture}}
\else{
\ifnum#1=1
{\begin{picture}(17,25)(1,-10)
\put(0,0){\circle{10}}
\put(0,0){\makebox(0,0)[c]{\scriptsize 2}}
\put(5,0){\line(1,0){5}}
\put(15,0){\circle{10}}
\put(15,0){\makebox(0,0)[c]{\scriptsize 1}}
\put(-7,0){\line(1,0){2}}
\put(20,0){\line(1,0){2}}
\put(0,5){\line(0,1){2}}
\put(15,5){\line(0,1){2}}
\put(0,-5){\line(0,-1){2}}
\put(15,-5){\line(0,-1){2}}
\end{picture}}
\else{
\ifnum#1=2
{\begin{picture}(31,10)(4,-10)
\put(0,0){\circle{10}}
\put(0,0){\makebox(0,0)[c]{\scriptsize 2}}
\put(5,0){\line(1,0){5}}
\put(15,0){\circle{10}}
\put(15,0){\makebox(0,0)[c]{\scriptsize 2}}
\put(20,0){\line(1,0){5}}
\put(30,0){\circle{10}}
\put(30,0){\makebox(0,0)[c]{\scriptsize 1}}
\put(-7,0){\line(1,0){2}}
\put(35,0){\line(1,0){2}}
\put(0,5){\line(0,1){2}}
\put(15,5){\line(0,1){2}}
\put(30,5){\line(0,1){2}}
\put(0,-5){\line(0,-1){2}}
\put(15,-5){\line(0,-1){2}}
\put(30,-5){\line(0,-1){2}}
\end{picture}}
\else{
\ifnum#1=3
{\begin{picture}(17,15)(10,-3)
\put(0,0){\circle{10}}
\put(0,0){\makebox(0,0)[c]{\scriptsize 2}}
\put(5,0){\line(1,0){5}}
\put(15,0){\circle{10}}
\put(15,0){\makebox(0,0)[c]{\scriptsize 2}}
\put(15,5){\line(0,1){5}}
\put(15,15){\circle{10}}
\put(15,15){\makebox(0,0)[c]{\scriptsize 1}}
\put(-7,0){\line(1,0){2}}
\put(20,0){\line(1,0){2}}
\put(8,15){\line(1,0){2}}
\put(20,15){\line(1,0){2}}
\put(0,-5){\line(0,-1){2}}
\put(0,5){\line(0,1){2}}
\put(15,20){\line(0,1){2}}
\put(15,-5){\line(0,-1){2}}
\end{picture}}
\else{
\ifnum#1=4
{\begin{picture}(40,10)(10,-3)
\put(0,0){\circle{10}}
\put(0,0){\makebox(0,0)[c]{\scriptsize 2}}
\put(5,0){\line(1,0){5}}
\put(15,0){\circle{10}}
\put(15,0){\makebox(0,0)[c]{\scriptsize 2}}
\put(20,0){\line(1,0){5}}
\put(30,0){\circle{10}}
\put(30,0){\makebox(0,0)[c]{\scriptsize 2}}
\put(35,0){\line(1,0){5}}
\put(45,0){\circle{10}}
\put(45,0){\makebox(0,0)[c]{\scriptsize 1}}
\put(-7,0){\line(1,0){2}}
\put(50,0){\line(1,0){2}}
\put(0,5){\line(0,1){2}}
\put(15,5){\line(0,1){2}}
\put(30,5){\line(0,1){2}}
\put(45,5){\line(0,1){2}}
\put(0,-5){\line(0,-1){2}}
\put(15,-5){\line(0,-1){2}}
\put(30,-5){\line(0,-1){2}}
\put(45,-5){\line(0,-1){2}}
\end{picture}}
\else{
\ifnum#1=5
{\begin{picture}(22,19)(15,5)
\put(0,0){\circle{10}}
\put(0,0){\makebox(0,0)[c]{\scriptsize 2}}
\put(5,0){\line(1,0){5}}
\put(15,0){\circle{10}}
\put(15,0){\makebox(0,0)[c]{\scriptsize 2}}
\put(20,0){\line(1,0){5}}
\put(30,0){\circle{10}}
\put(30,0){\makebox(0,0)[c]{\scriptsize 2}}
\put(30,5){\line(0,1){5}}
\put(30,15){\circle{10}}
\put(30,15){\makebox(0,0)[c]{\scriptsize 1}}
\put(-7,0){\line(1,0){2}}
\put(35,0){\line(1,0){2}}
\put(25,15){\line(-1,0){2}}
\put(35,15){\line(1,0){2}}
\put(0,5){\line(0,1){2}}
\put(15,5){\line(0,1){2}}
\put(30,20){\line(0,1){2}}
\put(0,-5){\line(0,-1){2}}
\put(15,-5){\line(0,-1){2}}
\put(30,-5){\line(0,-1){2}}
\end{picture}}
\else{
\ifnum#1=6
{\begin{picture}(31,15)(15,-3)
\put(0,0){\circle{10}}
\put(0,0){\makebox(0,0)[c]{\scriptsize 2}}
\put(5,0){\line(1,0){5}}
\put(15,0){\circle{10}}
\put(15,0){\makebox(0,0)[c]{\scriptsize 2}}
\put(15,5){\line(0,1){5}}
\put(15,15){\circle{10}}
\put(15,15){\makebox(0,0)[c]{\scriptsize 2}}
\put(20,15){\line(1,0){5}}
\put(30,15){\circle{10}}
\put(30,15){\makebox(0,0)[c]{\scriptsize 1}}
\put(-7,0){\line(1,0){2}}
\put(20,0){\line(1,0){2}}
\put(8,15){\line(1,0){2}}
\put(35,15){\line(1,0){2}}
\put(30,10){\line(0,-1){2}}
\put(30,20){\line(0,1){2}}
\put(0,-5){\line(0,-1){2}}
\put(0,5){\line(0,1){2}}
\put(15,20){\line(0,1){2}}
\put(15,-5){\line(0,-1){2}}
\end{picture}}
\else{
\ifnum#1=7
{\begin{picture}(25,35)(10,10)
\put(0,0){\circle{10}}
\put(0,0){\makebox(0,0)[c]{\scriptsize 2}}
\put(5,0){\line(1,0){5}}
\put(15,0){\circle{10}}
\put(15,0){\makebox(0,0)[c]{\scriptsize 2}}
\put(15,5){\line(0,1){5}}
\put(15,15){\circle{10}}
\put(15,15){\makebox(0,0)[c]{\scriptsize 2}}
\put(10,15){\line(-1,0){2}}
\put(20,15){\line(1,0){2}}
\put(15,30){\circle{10}}
\put(15,30){\makebox(0,0)[c]{\scriptsize 1}}
\put(-7,0){\line(1,0){2}}
\put(20,0){\line(1,0){2}}
\put(20,15){\line(1,0){2}}
\put(20,30){\line(1,0){2}}
\put(10,30){\line(-1,0){2}}
\put(15,35){\line(0,1){2}}
\put(0,-5){\line(0,-1){2}}
\put(0,5){\line(0,1){2}}
\put(15,20){\line(0,1){5}}
\put(15,-5){\line(0,-1){2}}
\end{picture}}
\else{
\ifnum#1=8
{\begin{picture}(15,40)(17,-2)
\put(0,0){\circle{10}}
\put(0,0){\makebox(0,0)[c]{\scriptsize 2}}
\put(5,0){\line(1,0){5}}
\put(15,0){\circle{10}}
\put(15,0){\makebox(0,0)[c]{\scriptsize 3}}
\put(15,5){\line(0,1){5}}
\put(15,15){\circle{10}}
\put(15,15){\makebox(0,0)[c]{\scriptsize 1}}
\put(15,-5){\line(0,-1){5}}
\put(15,-15){\circle{10}}
\put(15,-15){\makebox(0,0)[c]{\scriptsize 1}}
\put(-7,0){\line(1,0){2}}
\put(20,0){\line(1,0){2}}
\put(8,15){\line(1,0){2}}
\put(20,15){\line(1,0){2}}
\put(10,-15){\line(-1,0){2}}
\put(20,-15){\line(1,0){2}}
\put(0,-5){\line(0,-1){2}}
\put(0,5){\line(0,1){2}}
\put(15,20){\line(0,1){2}}
\put(15,-20){\line(0,-1){2}}
\end{picture}}
\else{
\ifnum#1=9
{\begin{picture}(2,15)(19,-3)
\put(0,0){\circle{10}}
\put(0,0){\makebox(0,0)[c]{\scriptsize 2}}
\put(5,0){\line(1,0){5}}
\put(15,0){\circle{10}}
\put(15,0){\makebox(0,0)[c]{\scriptsize 3}}
\put(20,0){\line(1,0){5}}
\put(30,0){\circle{10}}
\put(30,0){\makebox(0,0)[c]{\scriptsize 1}}
\put(15,5){\line(0,1){5}}
\put(15,15){\circle{10}}
\put(15,15){\makebox(0,0)[c]{\scriptsize 1}}
\put(-7,0){\line(1,0){2}}
\put(35,0){\line(1,0){2}}
\put(10,15){\line(-1,0){2}}
\put(20,15){\line(1,0){2}}
\put(0,5){\line(0,1){2}}
\put(15,20){\line(0,1){2}}
\put(30,5){\line(0,1){2}}
\put(0,-5){\line(0,-1){2}}
\put(15,-5){\line(0,-1){2}}
\put(30,-5){\line(0,-1){2}}
\end{picture}}
\else{
\ifnum#1=10
{\begin{picture}(30,25)(0,-3)
\put(0,0){\circle{10}}
\put(0,0){\makebox(0,0)[c]{\scriptsize 2}}
\put(5,0){\line(1,0){5}}
\put(15,0){\circle{10}}
\put(15,0){\makebox(0,0)[c]{\scriptsize 3}}
\put(20,0){\line(1,0){5}}
\put(30,0){\circle{10}}
\put(30,0){\makebox(0,0)[c]{\scriptsize 2}}
\put(15,5){\line(0,1){5}}
\put(15,15){\circle{10}}
\put(15,15){\makebox(0,0)[c]{\scriptsize 2}}
\put(30,15){\circle{10}}
\put(30,15){\makebox(0,0)[c]{\scriptsize 2}}
\put(-7,0){\line(1,0){2}}
\put(35,0){\line(1,0){2}}
\put(10,15){\line(-1,0){2}}
\put(20,15){\line(1,0){5}}
\put(35,15){\line(1,0){2}}
\put(0,5){\line(0,1){2}}
\put(15,20){\line(0,1){2}}
\put(30,5){\line(0,1){5}}
\put(30,20){\line(0,1){2}}
\put(0,-5){\line(0,-1){2}}
\put(15,-5){\line(0,-1){2}}
\put(30,-5){\line(0,-1){2}}
\end{picture}}
\else{
\ifnum#1=11
{\begin{picture}(5,20)(-5,-3)
\put(0,0){\circle{10}}
\put(0,0){\makebox(0,0)[c]{\scriptsize 2}}
\put(5,0){\line(1,0){5}}
\put(15,0){\circle{10}}
\put(15,0){\makebox(0,0)[c]{\scriptsize 3}}
\put(20,0){\line(1,0){5}}
\put(30,0){\circle{10}}
\put(30,0){\makebox(0,0)[c]{\scriptsize 2}}
\put(15,5){\line(0,1){5}}
\put(15,15){\circle{10}}
\put(15,15){\makebox(0,0)[c]{\scriptsize 1}}
\put(30,15){\circle{10}}
\put(30,15){\makebox(0,0)[c]{\scriptsize 3}}
\put(45,15){\circle{10}}
\put(45,15){\makebox(0,0)[c]{\scriptsize 1}}
\put(-7,0){\line(1,0){2}}
\put(35,0){\line(1,0){2}}
\put(10,15){\line(-1,0){2}}
\put(20,15){\line(1,0){5}}
\put(35,15){\line(1,0){5}}
\put(50,15){\line(1,0){2}}
\put(0,5){\line(0,1){2}}
\put(15,20){\line(0,1){2}}
\put(30,5){\line(0,1){5}}
\put(30,20){\line(0,1){2}}
\put(45,20){\line(0,1){2}}
\put(45,10){\line(0,-1){2}}
\put(0,-5){\line(0,-1){2}}
\put(15,-5){\line(0,-1){2}}
\put(30,-5){\line(0,-1){2}}
\end{picture}}
\else{
\ifnum#1=12
{\begin{picture}(5,10)(30,-3)
\put(0,0){\circle{10}}
\put(0,0){\makebox(0,0)[c]{\scriptsize 1}}
\put(5,0){\line(1,0){5}}
\put(15,0){\circle{10}}
\put(20,0){\line(1,0){5}}
\put(30,0){\circle{10}}
\put(30,0){\makebox(0,0)[c]{\scriptsize 1}}
\put(-7,0){\line(1,0){2}}
\put(35,0){\line(1,0){2}}
\put(0,5){\line(0,1){2}}
\put(15,5){\line(0,1){2}}
\put(30,5){\line(0,1){2}}
\put(0,-5){\line(0,-1){2}}
\put(15,-5){\line(0,-1){2}}
\put(30,-5){\line(0,-1){2}}
\end{picture}}
\else{
\ifnum#1=13
{\begin{picture}(-5,30)(30,-3)
\put(0,0){\circle{10}}
\put(0,0){\makebox(0,0)[c]{\scriptsize 1}}
\put(5,0){\line(1,0){5}}
\put(15,0){\circle{10}}
\put(15,5){\line(0,1){5}}
\put(15,15){\circle{10}}
\put(15,15){\makebox(0,0)[c]{\scriptsize 1}}
\put(0,15){\circle{10}}
\put(-7,0){\line(1,0){2}}
\put(-7,15){\line(1,0){2}}
\put(20,0){\line(1,0){2}}
\put(5,15){\line(1,0){5}}
\put(20,15){\line(1,0){2}}
\put(0,-5){\line(0,-1){2}}
\put(0,5){\line(0,1){5}}
\put(0,20){\line(0,1){2}}
\put(15,20){\line(0,1){2}}
\put(15,-5){\line(0,-1){2}}
\end{picture}}
\else
{\begin{picture}(0,10)(-3,-3)
\put(0,0){\circle{10}}
\put(0,0){\makebox(0,0)[c]{\scriptsize 2}}
\put(-7,0){\line(1,0){2}}
\put(5,0){\line(1,0){2}}
\put(0,5){\line(0,1){2}}
\put(0,-5){\line(0,-1){2}}
\end{picture}}
\fi}\fi}\fi}\fi}\fi}\fi}\fi}\fi}\fi}\fi}\fi}\fi}\fi}\fi}

\usepackage{enumitem}

\catcode`\@=11
\def\numberbysection{\@addtoreset{equation}{section}
\def\theequation{\arabic{section}.\arabic{equation}}}
\numberbysection

\def\be{\begin{equation}}
\def\ee{\end{equation}}
\newcommand\bea{\begin{eqnarray}}
\newcommand\eea{\end{eqnarray}}
\renewcommand\phi{\varphi}

\def\benn{\begin{eqnarray*}}
\def\eenn{\end{eqnarray*}}
\def\half{{\textstyle {1 \over 2}}}
\def\la{\langle}
\def\ra{\rangle}
\def\Z{{\mathbb Z}}
\def\Q{{\mathbb Q}}
\def\N{{\mathbb N}}
\def\P{{\mathbb P}}

\def\C{{\cal C}}
\def\L{{\cal L}}
\def\K{{\cal K}}
\def\R{{\cal R}}
\def\V{{\cal V}}

\def\half{{\textstyle {1\over 2}}}

\def\ci{{\rm i}}
\def\a{\alpha}
\def\b{\beta}

\def\bi{{\mathbb I}}

\renewcommand\tableofcontents{%
  \section*{\contentsname}%
  \@starttoc{toc}%
}
\catcode`@=12

\begin{document}

\hfill\today
\vspace{-10mm}

\title[The Abelian Sandpile Model]{Logarithmic conformal invariance in the Abelian sandpile model}

\author{Philippe Ruelle}
\address{Institute for Research in Mathematics and Physics, Universit\'e catholique de Louvain, B-1348 Louvain-la-Neuve, Belgium}
\begin{abstract}
We review the status of the two-dimensional Abelian sandpile model as a strong candidate to provide a lattice realization of logarithmic conformal invariance with central charge $c=-2$. Evidence supporting this view is collected from various aspects of the model. These include the study of some conformally invariant boundary conditions, and the corresponding boundary condition changing fields, the calculation of correlations of certain bulk and boundary observables (the height variables) as well as a proper account of the necessary dissipation, which allows for a physical understanding of some of the strange but generic features of logarithmic theories. 
\end{abstract}



\tableofcontents

\section{Introduction}

The Abelian sandpile model was originally conceived as a prototypical example of a so-called self-organized critical system. Namely, a dynamical system which automatically adjusts its own behaviour to converge and maintain itself in a critical state. According to its inventors, Bak, Tang and Wiesenfeld \cite{btw87}, the sandpile model is just a specific incarnation, in a concrete situation, of a generic mechanism present in Nature, which would explain why so many power laws are actually observed (earthquakes, avalanches, solar flares, ...).

Although criticality was built in the model from the start, the focus was rather on dynamical aspects and the extent to which the dynamics defining the sandpile model is universal. It was also soon realized that the asymptotic regime is described by a critical measure, and therefore lends itself to the usual analyses, like scaling and conformal invariance, especially in two dimensions. The first hint of the central charge ($c=-2$) as well as the non-local features of the model led to the suspicion that the conformal description could very well be logarithmic. 

If the abstract understanding of the general features of logarithmic theories has greatly improved, as this Special Issue clearly shows, the situation is not yet satisfactory in most cases. In this regard, lattice models realizing logarithmic conformal invariance may prove extremely useful. Infinitely many examples are known by now, some being better understood than others, but for most of them, it turns to be extremely difficult to completely pinpoint the specific conformal theory at work, a rather unusual situation if we compare with the unitary conformal theories. 

This review will focus exclusively on one of these lattice models, the two-dimensional Abelian sandpile model (ASM). In most lattice realizations, the conformal spectrum is more easily accessible while correlators are harder, and in many cases, unknown.  The ASM is one of the very few models where the situation is the inverse: correlation functions of some observables can be computed exactly, in the bulk, on boundaries, or both, whereas the complete conformal spectrum is not easy to determine because a transfer matrix formulation is not always available. This peculiarity makes the ASM rather unique among the lattice realizations of logarithmic conformal invariance, and helps to develop a physical understanding of this class of conformal theories.

We will start by giving a general overview of the ASM, as a discrete dynamical system in $2+1$ dimensions, and its relationship to other models (spanning trees, dimers, loop-erased random walks). In the asymptotic, stationary regime, the statistical properties of the model are controlled by a probability measure ${\mathbb P}$ on the space of configurations of the sandpile. For a finite system, this measure is simple but non-local with respect to the natural degrees of freedom, namely the heights of the pile at each site, which are strongly correlated throughout the lattice. Such intrinsically non-local features are a landmark of LCFTs, and qualifies the ASM as a good candidate for an LCFT description. Indeed most explicit computations to date aim at providing evidence that the scaling limit of the measure ${\mathbb P}$ is the field-theoretical measure of an LCFT, with central charge $c=-2$.

The rest of the article will be devoted to review the exact results which are the most convincing to support the previous assertion. Such calculations include the calculation of certain bulk correlators, the discussion of a few boundary conditions and the corresponding boundary condition changing fields, and some boundary and bulk correlators, some of them forming Jordan cells. In the ASM, an absolutely crucial r\^ole is played by dissipation, which is essential to make the dynamics well-defined. In the conformal picture, the insertion of dissipation is represented by a dimension 0 field, logarithmic partner of the identity, which allows for a transparent understanding of some of the strangest features of LCFT. 

We should mention that the ASM is not the only lattice model believed to be described by a logarithmic conformal theory with $c=-2$. At least two other models are known, namely the dense polymer model which, in its loop formulation, is the first of the infinite series of so-called logarithmic minimal models \cite{pearce2006}, and the dimer model \cite{izmailian2005}. In their scaling limit, the three models are believed to be different. Although very close and even equivalent in certain instances, the ASM and the dimer model are not when the lattice has one or more periodic direction (see below in Section \ref{related}). Some differences between the dense polymer model and the dimer model have been recently emphasized in \cite{rasmussen2012}. Finally, while the dense polymer is likely to be described by the symplectic free fermion theory, it is not the case of the ASM, as an argument recalled in Section \ref{uhp} shows. Exactly which conformal theories describe the dimer model and the ASM is a widely open question.


\section{The Abelian sandpile model}
\label{asm}

We review in this Section the definition of the model and its basic features, omitting most of the time the detailed proofs. These and more details about the model can be found f.i. in the reviews \cite{dhar2006,redig2005}. We also point out its relation to other mathematical problems. 

\subsection{Definition}

The sandpile model we are going to consider has been defined by Bak, Tang and Wiesenfeld in 1987 \cite{btw87}, but the first systematic mathematical analysis has been carried out by Dhar \cite{dhar1990}, who showed in particular its Abelian structure. The BTW model was subsequently called the Abelian sandpile model (ASM). The model was originally defined on a finite portion of $\Z^d$, but can as easily be defined on any finite graph $\L$, chosen unoriented to avoid technical subtleties \cite{speer1993}. As we will be primarily concerned with two-dimensional lattices, we will mostly think of $\L$ as being a finite portion $\L$ of $\Z^2$, typically a rectangular $L \times M$ grid, with $N = LM$ sites. The formulation and the results which follow are however general.

It is convenient to consider an extended graph $\L^\star$, which is the graph $\L$ itself, supplemented by one extra vertex called the {\it root} and edges connecting the root to sites of $\L$. For each vertex $i$, we accordingly define $z^{}_i$ as the coordination number of $i$ in $\L$ and $z_i^\star$ as its coordination number in $\L^\star$, so that $z_i^\star - z^{}_i \geq 0$ is the multiplicity of the edge connecting $i$ to the root. We also introduce the symmetric {\it toppling matrix} $\Delta$ on $\L$ as follows,
\be
\Delta_{i,j} = \cases{z_i^\star & for $i=j$,\cr
-1 & if $i$ and $j$ are neighbours,\cr
0 & otherwise.}
\ee

The ASM is a stochastic dynamical open system, discrete in space and in time. The microscopic degrees of freedom are random variables $h_i$, called heights, attached to the sites: $h_i$ counts the number of sand grains at site $i$ and takes integer values, conventionally chosen to be larger or equal to 1. A configuration is a set of values $\{h_i\}_{i \in \L}$; it is called {\it stable} if all heights satisfy $1 \leq h_i \leq z_i^\star$. We denote by $\cal S$ the set of all stable configurations, of cardinal $|{\cal S}| = \prod_{i \in \L} z^\star_i$.

The dynamics is defined on $\cal S$ as follows. Given a configuration $\C_t$ at time $t$, the dynamics produces the configuration $\C_{t+1}$ in two steps.
\begin{enumerate}[leftmargin=1.3cm]
\item[(a)] {\it Seeding}. A grain is dropped on a random site $j$ according to a distribution $\{p_j\}$ on $\L$, that is, $h_j \to h_j + 1$, and $h_i \to h_i$ for all $i \neq j$. $\C_{t+1}$ is this new configuration if it is stable; if it is not stable (new $h_j = z_j^\star + 1$), we go to step (b).
\item[(b)] {\it Relaxation}. Every unstable site --with a height larger or equal to $z_j^\star + 1$-- topples: it loses $z_j^\star$ grains of sand and gives one to each of its $z_j$ nearest neighbours in $\L$. In other words when a site $j$ topples, the whole configuration is updated as $h_i \to h_i - \Delta_{i,j}$. When all unstable sites have toppled, the resulting configuration is stable and defines $\C_{t+1}$. 
\end{enumerate}

We note that after a toppling at $j$, the sand is redistributed but remains within $\L$ if $j$ is not connected to the root ($z_j^\star = z_j^{}$); if $j$ is connected to the root, $z_j^\star - z_j^{}$ grains are transfered to the root and never come back in $\L$. In the former case, we say that the site $j$ is {\it conservative}, while in the latter case, it is called {\it dissipative}.

The above dynamics is well-defined. The seeding of a grain at $j$ can potentially trigger a large avalanche of topplings. If unstable after step (a), the site $j$ topples and transfers sand to its neighbours, which can themselves become unstable and topple, making their own neighbours unstable, and so on. However the relaxation process always terminates provided the set of dissipative sites is not empty and every site is path-connected to at least one dissipative site. Indeed if the avalanche were to propagate for ever, an infinite number of grains would be evacuated to the root, which is impossible since any stable configuration holds a finite quantity of sand. Thus the root acts like a infinitely deep {\it sink} to which sand is irreversibly lost. The root is therefore crucial to make the dynamics well-defined\footnote{This would not be the case if the model was defined from the start on an infinite lattice. The set of allowed configurations must then be appropriately restricted \cite{redig2005}.}.

In the cases that are most usually considered, namely a rectangular grid or a cylindrical grid, the dissipative sites are chosen among the boundary sites. A boundary site connected to the root is called {\it open}, and {\it closed} if it is not connected to the root. Similarly a whole boundary consisting of open resp. closed sites is called an open resp. closed boundary. The bulk sites are usually not connected to the root, and so have $z^{}_i = z_i^\star = 4$ on the square lattice. The toppling matrix becomes a Laplacian-like operator with appropriate boundary conditions.

The relaxation process is also independent of the order in which the topplings take place. This is a consequence of the fact that whether or not a toppling occurs at site $j$ depends only on $h_j$, and that the effects of two topplings at $j$ and $k$ commute,
\be
h_i \to h_i - \Delta_{i,j} - \Delta_{i,k} = h_i - \Delta_{i,k} - \Delta_{i,j}.
\ee

If we denote by $a_j$ the operator that drops a sand grain at $j$ and executes the relaxation process, we can rephrase the dynamics by saying that given $\C_t$, it returns $\C_{t+1} = a_j \, \C_t$ with probability $p_j$. The operators $a_j$, for $j \in \L$, all map the set of stable configurations into itself, and from the previous observations, are mutually commuting,
\be
[a_i , a_j] = 0, \qquad \forall i,j \in \L.
\ee
The qualification of this sandpile model as Abelian stems from this property.

The dynamics as defined above allows to follow the stochastic time evolution of any given initial configuration. As we are interested in the long-time statistical properties of the configurations $\C_t$, it is more convenient to compute the time evolution of probability distributions $\P_t(\C)$ over the set of stable configurations. From an initial distribution $\P_0(\C)$, one easily obtains the following master equation,
\be
\P_{t}(\C) = \sum_{j \in \L} \: p_j \, \sum_{\C'} \: \delta(\C - a_j \, \C') \, \P_{t-1}(\C') = \sum_{\C'} \: W_{\C,\C'} \, \P_{t-1}(\C').
\ee
The matrix $W_{\C,\C'}$ is the transition matrix for a Markov chain with finite state space $\cal S$,  equivalently of a random walk on $\cal S$.

The long-time behaviour of the sandpile is controlled by the limit measure $\P^*_\L = \lim_{t \to \infty} \, \P_t = \lim_{t \to \infty} \, W^t \, \P_0$. The general theory of Markov chains applied to this particular case yields the following results.  

The set of stable configurations is divided into two disjoint subsets: the {\it recurrent} configurations are in the repeated image of the dynamics and keep reoccuring infinitely often; those which can only occur a finite number of times are called {\it transient}. For all $t$ large enough, $\P_t(\C) = 0$ if $\C$ is transient and $\P_t(\C) > 0$ if $\C$ is recurrent. Therefore if we start from any stable configuration and apply the dynamics, it takes a finite time (of order $N^2$) to reach a recurrent configuration, after which all subsequent configurations are recurrent (from the definitions, a recurrent configuration can never be mapped to a transient one).

Under the hypothesis that $p_j > 0$ for all $j \in \L$ (the uniform distribution $p_j = {1 \over N}$ is usually used), one can show that the limit $\P^*_\L$ is unique, i.e. does not depend on the initial distribution, and moreover is uniform on the set $\cal R$ of recurrent configurations,
\be
\P^*_\L(\C) = \cases{{1 \over |{\cal R}|} & if $\C$ is recurrent,\cr
0 & if $\C$ is transient.}
\ee

The stationary distribution $\P^*_\L$ for a given graph looks very simple, since it is uniform, but at the same time is extremely complex because its support, the set $\cal R$, has a complicated structure, as we will see below. In fact, a configuration must satisfy non-local contraints to be recurrent. 

In the infinite volume limit, the sequence of measures $\P^*_\L$ converges to a measure on an infinite lattice, which we simply denote by $\P$. The main purpose of this review is to give support to the  statement that

\noindent
\vrule width 0.5mm depth 1.0cm
\vspace{-1.0cm}

\noindent
\leftskip=5mm
{\it The scaling limit of the measure $\P$ is the field-theoretic measure of a two-dimensional logarithmic conformal field theory with central charge $c=-2$.}

\vspace{4mm}\leftskip=0mm 
\noindent
Evidence will be collected from results obtained during the last ten years.

\subsection{Recurrent configurations and spanning trees}

The recurrent configurations manifestly play a central r\^ole. The few definitions recalled above allow us to gain some insight into what they really are. We assume all $p_i > 0$.

A first observation is that the full (or maximal) configuration $\C^\star$ with $h_i =z_i^\star$ is certainly recurrent. Indeed it can be obtained from any stable configuration by applying a suitable sequence of $a_i$ (by filling the sites which are strictly below their threshold). Then any configuration of the form $\prod_j a_j^{k_j}\C^\star$ obtained from $\C^\star$ by applying any string of $a_j$ is also recurrent. Conversely, any recurrent configuration must be of this form, because once $\C^\star$ is produced by the dynamics, they are the only ones to ever appear. As any recurrent configuration can be obtained from $\C^\star$ and vice-versa, it follows that any recurrent configuration can be obtained from any other by applying a suitable string of $a_i$; they form under the dynamics a single irreducible component. Another simple corollary is that if $\C=\{h_i\}$ is recurrent, then any stable configuration $\C'$ with heights $h'_i \geq h_i$ is also recurrent. 

The previous observations imply that the operators $a_i$ are invertible on $\R$. For every $i$, an operator $b_i$, monomial in the $a$'s, should exist such that $b_ia_i\,\C^\star = a_ib_i\,\C^\star = \C^\star$. It must also be unique because the only recurrent configuration $\C_i$ such that $a_i\C_i = \C^\star$ has heights equal to $h^{}_k = z_k^\star - \delta_{k,i}$. Thus the operators $a_i$ restricted to the recurrent configurations have a unique inverse and generate a finite Abelian group.

The question nonetheless remains: are all stable configurations recurrent ? The negative answer stems from the toppling rules.

To take a simple example, consider a stable configuration in which two neighbouring sites have a height equal to 1. Assume also that the dynamics has been run for long enough so that these two sites have both toppled. The site that has toppled last has given one grain to its neighbour which therefore must have a height at least equal to 2. Thus any stable configuration with two neighbouring 1's cannot be recurrent. The same argument applies to three neighbouring sites with height values 121 (the 2 in central position) or to four neighbouring sites forming a square, all with a height 2.

More generally, a subconfiguration on a subset of sites $F$ is forbidden in a recurrent configuration if each site of $F$ has a height which is smaller or equal to the number of its neighbours in $F$. The three examples given above are the smallest forbidden clusters; however the size of a forbidden cluster is only bounded by the size of $\L$. For instance the configuration $h_i = z_i$ on $\L$ is forbidden for $F = \L$ and contains no smaller forbidden subconfiguration. 

Forbidden subconfigurations form the key concept because one can prove that {\it a configuration is recurrent if and only if it contains no forbidden subconfiguration} \cite{majdhar1992}. Because one may have to scan a large portion of the lattice to see if it contains a forbidden subconfiguration, the recurrence property is non-local. This is what makes the sandpile model non-trivial, interesting but also notoriously difficult to handle analytically\footnote{Notable exceptions are Bethe lattices \cite{dharmaj1990} or 1d-like lattices \cite{rusen1992,alidhar1995}, but admittedly they show less interesting behaviour.}. 

The {\it burning algorithm} \cite{dhar1990} provides a simple and practical way to test whether a configuration is recurrent and at the same time establishes \cite{majdhar1992} a bijective correspondence with rooted spanning trees on $\L^\star$ (hence an equivalence of the sandpile model and the $q \to 0$ limit of the $q$-Potts model). It goes like follows.

To start, let $\K_0=\L$ be the set of unburnt sites at time 0. The sites of $\K_0$ whose height is strictly larger than their number of neighbours in $\K_0$ are called burnable at time 1 (these sites must be dissipative). We burn them, and doing so we define a new set of unburnt sites $\K_1 \subseteq \K_0$ at time 1. Next we burn the sites of $\K_1$ which are burnable at time 2 (those with a height larger than their number of neighbours on $\K_1$) and obtain yet a new set $\K_2 \subseteq \K_1$ of unburnt sites at time 2. This procedure is carried on until, for some $T$, no site of $\K_T$ is burnable, implying  $\K_{T+1} = \K_T$. If $\K_T = \{\emptyset\}$, the configuration is recurrent; otherwise, the subconfiguration on $\K_T$ is forbidden and the configuration is transient.

In the sequence of burnings as described above, it is not difficult to see that a site $j$ of $\K_t$ which is burnable at time $t+1$ has a least one neighbour that was burnt at time $t$ (otherwise $j$ would have been burnable at an earlier time). If $j$ has exactly one such neighbour $k$, we say that $j$ catches fire from $k$ or that the fire propagates along the bond $(k,j)$. If there are more than one, we use an ordering prescription to decide which neighbour $k$ sets $j$ afire. The prescription is arbitrary --and may change from site to site-- but it must be fixed, i.e. identical for all configurations. We also conventionally decide that the sites of $\K_0$ which are burnable at time 1 catch fire from the root. If the configuration is recurrent, the fire path is a collection of edges which contains no loop and such that every site is the endpoint of at least one edge. In other words, the fire path forms a {\it rooted spanning tree}, growing from the root towards the interior of $\L$. Conversely, and given a fixed ordering prescription, every spanning tree is associated with a unique recurrent configuration. Therefore the measure induced on the spanning trees is also uniform.

Note that the rooted spanning tree one obtains from this algorithm is connected on $\L^\star$. If one removes the root, and therefore focus on $\L$, the tree in general will appear to be disconnected, i.e. will form a spanning forest. 

We can observe that the rooted spanning tree carries a natural orientation, defined by the fire propagation. It is however conventional to orient the rooted spanning tree in the opposite direction, so that the oriented edges are like arrows that flow towards the root. A recurrent configuration can then be viewed as a configuration of arrows: each site has one outgoing arrow pointing to one of its neighbours in $\L^\star$ (it can be the root for dissipative sites) with the global constraint that the arrows do not form loops. 

The total number of arrow configurations is clearly equal to $\prod_{i \in \L}\, z_i^\star = |{\cal S}|$, but the number of those without loops is smaller. By the matrix-tree (Kirchhoff) theorem, this number, also equal to the number of recurrent configurations, is equal to
\be
|\R| = \det \, \Delta.
\ee
We note that if $\L$ contains no dissipative site, that is $z_i^\star = z_i^{}$ for all $i$, $\det \Delta$ vanishes and no configuration is recurrent (there are no spanning trees).

For $\L$ a thick domain in $\Z^2$ containing $N$ sites in which the interior sites are conservative, one can show that the number of recurrent configurations on $\L$ grows exponentially like $|\R| \simeq {\rm e}^{{4{\rm G} \over \pi}N} \simeq (3.21)^N$ with ${\rm G} =  0.915965$ the Catalan contant, to be compared with the number $\simeq 4^N$ of all stable configurations. 

Thus the ASM configurations can be described in terms of height variables or spanning trees. The heights may look more natural, they are local but globally constrained. The global constraints induced by the recurrence property are encoded into the global structure of the spanning trees. In actual computations, the latter often prove more convenient.

\subsection{The sandpile group}

The ASM offers a large number of interesting and challenging mathematical problems \cite{levine2010}. A fine example (see Section \ref{local} for others) is the sandpile group, also called the critical group of a graph. 

We have shown that the operators $a_i$ generate a finite Abelian group $G$ (it depends on $\L$ but we omit the explicit dependence). The order of $G$ is equal to $|\R|$, because all recurrent configurations can be obtained from any fixed one by repeated applications of the $a_i$. The recurrent configurations themselves form an Abelian group isomorphic to $G$, with a group law given by the sitewise addition of heights (followed by relaxation)
\be
\C \oplus \C' = \prod_{i \in \L} \: a_i^{h^{}_i} \, \C' = \prod_{i \in \L} \: a_i^{h'_i} \, \C\,.
\ee
In particular, there must be a unique recurrent configuration which is the identity in $G$. The resolution of $G$ into cyclic factors and the geometric structure of the identity configuration are challenging questions. 

The addition of $\Delta_{i,i}=z^\star_i$ grains at site $i$ of a recurrent configuration makes it topple. After the toppling of $i$, the height $h_i$ returns to its original value while the neighbouring heights have been incremented by 1. Therefore adding $\Delta_{i,i}$ grains at $i$ is the same as adding one grain to each neighbour of $i$. This yields the following relations on the generators of $G$,
\be
\prod_{j \in \L} \: a_j^{\Delta{i,j}} = 1, \qquad \forall i \in \L.
\label{rel}
\ee

Being finite Abelian, the irreducible representations of $G$ are all one-dimensional, given by $a_j = {\rm e}^{2{\rm i}\pi \phi_j}$. The relations (\ref{rel}) imply that $\sum_j \Delta_{i,j}\phi_j = m_i$ for an integer vector $\vec m \in \Z^N$, and so $\phi_j = \sum_{k \in \L} \, (\Delta)^{-1}_{j,k} \, m_k$ are rational. Since two vectors $\vec m$ and $\vec m + \Delta \vec n$ yield identical representations (same phases $\phi_{j}$), the dual group of $G$ and thus $G$ itself is given by
\be
G = \Z^N / \Delta \, \Z^N.
\label{Gr}
\ee
In particular we recover the order $|G| = \det\Delta$. 

In terms of recurrent configurations, this last equation has an interesting consequence. Let us consider the set of all configurations with heights $h_i \in \Z$ and the equivalence $\C \sim \C'$ if and only if $\C$ and $\C'$ are related by topplings, i.e. $\C' = \C + \Delta \vec n$ for some integer $\vec n$. Then the quotient $\Z^N / \Delta\Z^N$ is the number of equivalence classes, and (\ref{Gr}) shows that there is exactly one recurrent configuration in each class. In this respect, toppling invariant functions on $\R$ are particularly useful as coordinates on $G$.

The group structure of $G$ is identical to $\Z^N / \Delta \, \Z^N$ as a finite $\Z$-module. On general grounds \cite{jacobs1974}, it is a direct product of finite cyclic groups
\be
G = \Z_{d_1} \times \Z_{d_2} \times \ldots \times \Z_{d_N}, \qquad d_{i+1} | d_i.
\label{elem}
\ee
The integers $d_i$, called the elementary divisors of $\Delta$, are related to the Smith normal form\footnote{If $\Delta$ is a non-singular matrix with entries in $\Z$, the Smith normal form of $\Delta$ is the unique diagonal matrix $D$ such that $\Delta = A D B$ for $A,B$ in GL$(N,\Z)$ and $D_{ij} = d_i \delta_{ij}$ for positive integers $d_i$ satisfying $d_{i+1} | d_i$ \cite{jacobs1974}.}  of $\Delta$ (generically, most of them are equal to 1). For a specific, small graph, the computation of the elementary divisors is standard, but to compute them for large graphs or infinite classes of graphs is highly non-trivial. The generators of the cyclic factors in (\ref{elem}) as well as a complete set of toppling invariants have been explicitly constructed using the Smith normal form of $\Delta$ \cite{dharco1995}.

Many works have been devoted to this problem. While some of them focus on general properties \cite{cori2000,toumpa2007}, others consider specific classes of graphs. When the graph is a square $L \times L$ grid in $\Z^2$, it has been shown in \cite{dharco1995}, by using toppling invariants, that $G$ has exactly $L$ non-trivial factors in (\ref{elem}). For $L=2,3,4,5$ for instance, one finds the following groups
\bea
G_{2 \times 2} &=& \Z_{24} \times \Z_{8}, \qquad G_{3 \times 3} = \Z_{224}\times \Z_{112} \times \Z_4,\\
G_{4 \times 4} &=& \Z_{6600} \times \Z_{1320} \times \Z_8 \times \Z_8,\\
G_{5 \times 5} &=& \Z_{102960} \times \Z_{102960} \times \Z_{48} \times \Z_{16} \times \Z_4.
\eea

The structure of $G$ has been completely determined for wheel graphs $W_n$ \cite{biggs1999}, complete graphs $K_n$ \cite{bjacob2003}, graphs of dihedral groups $D_n$ \cite{dartois2003}, square cyclic graphs \cite{houa2006}, and others, mainly Cartesian products of simple graphs. 

The identity configuration is easy to construct algorithmically. It is the only recurrent configuration in the equivalence class of the empty configuration $h_i=0$, or of any configuration such that $\prod_{i \in \L} a_i^{h_i} = 1$. A convenient choice is to start from $\C_0$ with heights $h_i = \sum_{j \in \L} \Delta_{i,j} = z_i^\star-z_i$. It is not recurrent though since the conservative sites have zero height. However the sequence $\C_n = \C_0 \oplus \C_{n-1}$ stabilizes for a finite index $k$, at which point $\C_k = \C_{k-1}$ is recurrent and provides the group identity \cite{creutz1991}. 

The identity configuration has been studied mainly for rectangular portions of $\Z^2$, where it shows intriguing and complicated fractal patterns, first explored in \cite{creutz1991}. Its geometric structure in general remains poorly understood. Partial answers have been obtained for certain rectangular domains \cite{leborgne2006}. For square grids, very little is known apart from the fact that the identity on the $(2L+1) \times (2L+1)$ grid is simply related to that on the $2L \times 2L$ grid \cite{dharco1995}. The identity has been fully determined on certain oriented square grids \cite{cara2008}.

\subsection{Related models}
\label{related}

Through the relation between recurrent configurations and rooted spanning trees, explained above, the stationary regime of the ASM is connected to two other models, the dimer model and the loop-erased random walk (LERW).

A dimer configuration (also known as a perfect matching in the mathematical literature) on a graph with an even number of vertices is a subset of edges such that every vertex belongs to exactly one edge. The dimer model then investigates the statistics of the dimer configurations. It is quite old as it was proposed in 1937 as a model of adsorption of diatomic molecules on a subtrate \cite{fowler1937}. Since then it has been the subject of a vast literature, especially in recent years, see \cite{kenyon2009} for a recent review.

The correspondence with spanning trees has been first observed by Temperley \cite{temperley1974} for odd-by-odd rectangles in $\Z^2$ with a corner removed. It is pictured in Figure 1. The dimers touching the odd sublattice (both coordinates are odd) are marked in blue, the others in red. Noting that the red dimers are unambigously fixed by the blue ones (or vice-versa), we may focus on the blue dimers. Replacing them by arrows, we obtain a spanning tree rooted at the removed corner, and living on the odd sublattice. 

This correspondence can be generalized to any rectangle and actually holds for any bipartite planar graph \cite{kenyon2000,kenyon2004}. If the dimers are uniformly weighted, the correspondence induces a uniform distribution on trees, which therefore connects to the uniform measure on the recurrent configurations. When the lattice has a cylindrical or toroidal geometry, spanning trees are to be complemented by cycle-rooted spanning forests \cite{brankov2013}.

\begin{figure}[t]
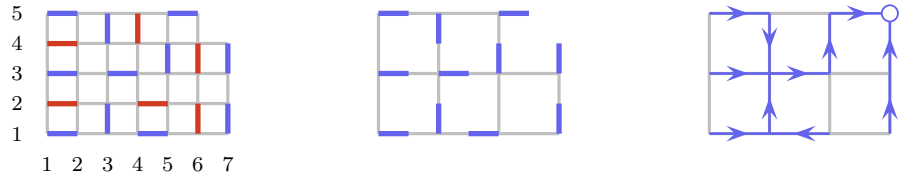

\psset{xunit=0.4cm}
\psset{yunit=0.4cm}
\psset{runit=0.4cm}
\def\hodimer{\psline[linewidth=2pt,linecolor=myblue](0,0)(1,0)}
\def\vodimer{\psline[linewidth=2pt,linecolor=myblue](0,0)(0,1)}
\def\hedimer{\psline[linewidth=2pt,linecolor=lightred](0,0)(1,0)}
\def\vedimer{\psline[linewidth=2pt,linecolor=lightred](0,0)(0,1)}
\hfill \pspicture(-1.5,-2)(7,5.5)
\multido{\nt=0+1}{4}{\psline[linewidth=1.2pt,linecolor=mygray](0,\nt)(6,\nt)}
\multido{\nt=0+1}{6}{\psline[linewidth=1.2pt,linecolor=mygray](\nt,0)(\nt,4)}
\psline[linewidth=1.2pt,linecolor=mygray](0,4)(5,4)
\psline[linewidth=1.2pt,linecolor=mygray](6,0)(6,3)
\rput(-1,-1){\multido{\nt=1+1}{5}{\rput(0,\nt){\scriptsize $\nt$}}}
\rput(-1,-1){\multido{\nt=1+1}{7}{\rput(\nt,0){\scriptsize $\nt$}}}
%
%
\rput(0,0){\hodimer}
\rput(2,0){\vodimer}
\rput(3,0){\hodimer}
\rput(5,0){\vedimer}
\rput(6,0){\vodimer}
\rput(0,1){\hedimer}
\rput(3,1){\hedimer}
%
\rput(0,2){\hodimer}
\rput(2,2){\hodimer}
\rput(4,2){\vodimer}
\rput(5,2){\vedimer}
\rput(6,2){\vodimer}
\rput(0,3){\hedimer}
\rput(2,3){\vodimer}
\rput(3,3){\vedimer}
%
\rput(0,4){\hodimer}
\rput(4,4){\hodimer}
\endpspicture
\hspace{1cm}
\pspicture(-1.5,-2)(7,5.5)
\multido{\nt=0+2}{2}{\psline[linewidth=1.2pt,linecolor=mygray](0,\nt)(6,\nt)}
\multido{\nt=0+2}{3}{\psline[linewidth=1.2pt,linecolor=mygray](\nt,0)(\nt,4)}
\psline[linewidth=1.2pt,linecolor=mygray](0,4)(5,4)
\psline[linewidth=1.2pt,linecolor=mygray](6,0)(6,3)
%
%
\rput(0,0){\hodimer}
\rput(2,0){\vodimer}
\rput(3,0){\hodimer}
\rput(6,0){\vodimer}
\rput(0,2){\hodimer}
\rput(2,2){\hodimer}
\rput(4,2){\vodimer}
\rput(6,2){\vodimer}
\rput(2,3){\vodimer}
\rput(0,4){\hodimer}
\rput(4,4){\hodimer}
\endpspicture
\hspace{1cm}
\pspicture(-1.5,-2)(7,5.5)
\def\br{\psline[linewidth=1.0pt,linecolor=myblue](1.,0)(2,0)
\psline[linewidth=1.0pt,linecolor=myblue,arrowsize=4pt 2]{->}(0,0)(1.3,0)}
\def\bl{\psline[linewidth=1.0pt,linecolor=myblue](-1,0)(-2,0)
\psline[linewidth=1.0pt,linecolor=myblue,arrowsize=4pt 2]{->}(0,0)(-1.2,0)}
\def\bu{\psline[linewidth=1.0pt,linecolor=myblue](0,1)(0,2)
\psline[linewidth=1.0pt,linecolor=myblue,arrowsize=4pt 2]{->}(0,0)(0,1.2)}
\def\bd{\psline[linewidth=1.0pt,linecolor=myblue](0,-1)(0,-2)
\psline[linewidth=1.0pt,linecolor=myblue,arrowsize=4pt 2]{->}(0,0)(0,-1.2)}
\multido{\nt=0+2}{2}{\psline[linewidth=1.2pt,linecolor=mygray](0,\nt)(6,\nt)}
\multido{\nt=0+2}{3}{\psline[linewidth=1.2pt,linecolor=mygray](\nt,0)(\nt,4)}
\psline[linewidth=1.2pt,linecolor=mygray](0,4)(5,4)
\psline[linewidth=1.2pt,linecolor=mygray](6,0)(6,3)
%
%
\rput(0,0){\br}
\rput(2,0){\bu}
\rput(4,0){\bl}
\rput(6,0){\bu}
\rput(0,2){\br}
\rput(2,2){\br}
\rput(4,2){\bu}
\rput(6,2){\bu}
\rput(0,4){\br}
\rput(2,4){\bd}
\rput(4,4){\br}
\pscircle[linewidth=0.7pt,linecolor=myblue,fillstyle=solid,fillcolor=white](6,4){0.3}
\endpspicture
\caption{Temperley's correspondence. The left figure shows a typical dimer configuration on an odd-by-odd rectangle with the upper right corner removed. In the central panel, the dimers touching the odd sublattice are kept, and replaced by arrows in the right figure. The arrows form a rooted spanning tree, where the root is represented by a circle. In ASM terms, there are only two dissipative sites.}
\end{figure}

The LERW has been proposed by Lawler \cite{lawler1980} as a simpler version of the self-avoiding random walks (SARW). The sample paths are defined from a symmetric random walks by chronologically removing the loops as they form (and simultaneously shifting the clock backwards so that a LERW at time $t$ has $t$ steps). In contrast, in the SARW, the finite paths are those of the symmetric random walk which do not cross themselves. The finite paths in the LERW and SARW are identical but their construction are endowed with different probability measures.

The relation between rooted spanning trees and LERW has been established in \cite{pemantle1991,majumdar1992}. The measure on the LERW sample paths connecting two points $i$ and $j$ coincides with the measure on chemical paths of uniform random spanning trees between $i$ and $j$. Wilson's algorithm \cite{wilson1996} to generate spanning trees with uniform measure is based on this connection. 

Until recently, the simple-looking question asking the passage probability through a given site or a given edge had remained open, but substantial progress has been made by Kenyon and Wilson \cite{kenyon2011} who actually computed such probabilities on $\Z^2$ (also on the honeycomb and triangular lattices). As we will see below, the LERW passage probabilities are directly related to height probabilities in the ASM, and lead to similarly new results for joint height probabilities.


\section{On representations}
\label{basics}

A conformal theory, logarithmic or not, is primarily specified by the type (and number) of representations it contains. In the logarithmic case however, the situation is notoriously and dramatically complicated because the sorts of reducible indecomposable representations that may potentially appear can be of considerable complexity. The only purpose of this section is to briefly review and set the notation for the simplest chiral representations at $c=-2$ that will be considered in the sequel, with no claim of exhaustivity (by far). Much more general results are reviewed elsewhere in this Special Issue \cite{lcft}. We should also mention that most of the works cited in this section generalize in various ways the results given below to other values of the central charge. 

Highest weight representations form the simplest class. A highest weight representation is generated from a single (highest weight) vector $|h\ra$ with conformal dimension $h$, i.e. satisfying $L_n |h\ra = (L_0 - h)|h\ra = 0$ for $n>0$. The full representation coincides with the Verma module $V_h$, obtained by acting with the negative Virasoro modes on $|h\ra$, and is irreducible for generic values of $h$. 

The module $V_h$ becomes reducible when $h$ belongs to the Kac table, which, for $c=-2$, is the following set of rational numbers,
\be
\hspace{-1.5cm}
\left\{h_{r,s} = {(2r-s)^2 - 1 \over 8} \;:\; r,s \leq 1 \right\} = \{\textstyle -{1 \over 8},\,{3 \over 8},\,{15 \over 8},\,{35 \over 8},\ldots\} \cup \{0,\,1,\,3,\,6,\ldots\}.
\ee
In particular $V_{r,s} \equiv V_{h_{r,s}}$ contains a singular vector at level $rs$ which generates a submodule $V_{r+s,s}$, and which allows to define the quotient representation
\be
{\cal V}_{r,s} = V_{r,s}/V_{r+s,s}.
\ee

The representations ${\cal V}_{r,s}$ are still highest weight and generally reducible indecomposable, except those for $s=1,2$ or $r=1$ and $s$ even, which are irreducible. They are however not closed under fusion and require to consider new representations, first studied in \cite{rohsiepe1996}, and shortly after in \cite{gaberdiel1996} where they were noted ${\cal R}_{r,1}$, $r \geq 1$. The representations ${\cal R}_{r,1}$ are not highest weight representations: they are reducible indecomposable and carry a non-diagonalizable action of $L_0$. 

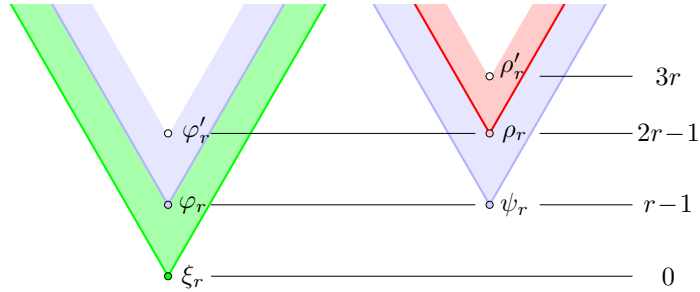
\begin{figure}[tbp]
\begin{center}
\psset{xunit=.95cm}
\psset{yunit=.95cm}
\psset{runit=.95cm}
\begin{pspicture}(0.5,2.4)(7,6.7)
\psclip{\psline[linestyle=none](-0.8,-0.)(-0.8,6.3)(9.5,6.3)(9.5,-0.)}
\pswedge[linecolor=green,fillstyle=solid,fillcolor=lightgreen](1.5,2.5){5}{60}{120}
\pswedge[linecolor=midblue,fillstyle=solid,fillcolor=lightblue2](1.5,3.5){5}{60}{120}
\pswedge[linecolor=white,fillstyle=solid,fillcolor=white](1.5,4.5){5}{60}{120}
\pscircle[fillstyle=solid,linewidth=0pt,fillcolor=green](1.5,2.5){0.05}
\pscircle[fillstyle=solid,linewidth=0pt,fillcolor=lightblue](1.5,3.5){0.05}
\pscircle[fillstyle=solid,linewidth=0pt,fillcolor=white](1.5,4.5){0.05}
\rput(1.85,2.5){\footnotesize $\xi_r$}
\rput(1.85,3.5){\footnotesize $\phi_r$}
\rput(1.9,4.55){\footnotesize $\phi'_r$}
\pswedge[linecolor=midblue,fillstyle=solid,fillcolor=lightblue2](6,3.5){5}{60}{120}
\pswedge[linecolor=red,fillstyle=solid,fillcolor=lightred2](6,4.5){5}{60}{120}
\pswedge[linecolor=white,fillstyle=solid,fillcolor=white](6,5.3){5}{60}{120}
\pscircle[fillstyle=solid,linewidth=0pt,fillcolor=lightblue](6,3.5){0.05}
\pscircle[fillstyle=solid,linewidth=0pt,fillcolor=lightred2](6,4.5){0.05}
\pscircle[fillstyle=solid,linewidth=0pt,fillcolor=white](6,5.3){0.05}
\rput(6.35,3.5){\footnotesize $\psi_r$}
\rput(6.35,4.5){\footnotesize $\rho_r$}
\rput(6.32,5.45){\footnotesize $\rho'_r$}
\psline[linecolor=black,linewidth=0.1pt](2.1,2.5)(8,2.5)
\psline[linecolor=black,linewidth=0.1pt](2.1,3.5)(5.8,3.5)
\psline[linecolor=black,linewidth=0.1pt](2.1,4.5)(5.8,4.5)
\psline[linecolor=black,linewidth=0.1pt](6.7,3.5)(8,3.5)
\psline[linecolor=black,linewidth=0.1pt](6.7,4.5)(8,4.5)
\psline[linecolor=black,linewidth=0.1pt](6.7,5.3)(8,5.3)
\rput(8.5,2.5){\footnotesize $0$}
\rput(8.5,3.5){\footnotesize $r\!-\!1$} 
\rput(8.5,4.5){\footnotesize $2r\!-\!1$}
\rput(8.5,5.3){\footnotesize $3r$}
\endpsclip
\end{pspicture}
\end{center}
\caption{Graphical presentation of the representations ${\cal R}_{r,1}$. The two cone-like figures picture the representations ${\cal V}_{1,2r-1}$ (left) and ${\cal V}_{1,2r+1}$ (right), where the single-colored shaded regions represent the irreducible subfactors. The states $\phi'_r$ and $\rho'_r$ are singular null states and generate the null Verma modules (in white). The numbers on the right show the levels at which the various states occur. For $r=1$, the proper figure is obtained by removing the green region (and the state $\xi_r$).}
\label{Rr1}
\end{figure}

${\cal R}_{r,1}$ is made up of two representations ${\cal V}_{1,2r-1}$ and ${\cal V}_{1,2r+1}$, whose highest weights differ by $r-1$, tied together by rank 2 Jordan cells. They are pictorially represented in Figure \ref{Rr1}. The states $\phi_r$ and $\phi'_r$ are the two lowest singular states in ${\cal V}_{1,2r-1}$, and only $\phi'_r$ is set to zero. The situation is similar in the right representation ${\cal V}_{1,2r+1}$ with the two states $\rho_r$ and $\rho'_r$.

The left representation ${\cal V}_{1,2r-1}$ is a subrepresentation while the right representation ${\cal V}_{1,2r+1}$ is not: the Virasoro modes have an off-diagonal action from ${\cal V}_{1,2r+1}$ to ${\cal V}_{1,2r-1}$. The main defining relations read
\bea
&& L_0 \psi_r = h_{1,2r+1}\psi_r + \phi_r, \quad L_1^{r-1} \psi_r = \xi_r, \quad L_n \psi_r = 0 \;\;{\rm for\ }n \geq 2,\\
&& (L_{-1}^{r-1} + \ldots) \,\xi_r = \beta_r \, \phi_r.
\eea
The pair $(\phi_r,\psi_r)$ is the lowest-lying Jordan cell in ${\cal R}_{r,1}$, and gives rise to an infinite number of descendant rank 2 Jordan blocks lying at higher levels.

An important feature of these representations is that the presence of the parameter $\beta_r$. Once $\psi_r$ is given, the fields (or the corresponding states) $\phi_r,\xi_r$ with their normalization are fixed. However $\phi_r$, as a singular descendant of $\xi_r$ at level $r-1$, must be equal to some multiple, say $\beta_r^{-1}$, of a canonical choice like $(L_{-1}^{r-1} + \ldots ) \, \xi_r$. The parameter $\beta_r$ cannot be absorbed in the normalizations of the various fields, and so its value is intrinsically related to the conformal transformation of $\psi_r$. It is therefore a number that labels the equivalence classes: two representations of type ${\cal R}_{r,1}$ are inequivalent if and only if they have different values of $\beta_r$ \cite{kytola2009} (though all representations for fixed $r$ have identical character). 

The parameter $\beta_r$ is known in the literature as a logarithmic coupling, an indecomposability parameter or a beta-invariant. A general mathematical study of chiral rank 2 staggered modules, of which the ${\cal R}_{r,1}$ are specific examples, has been carried out in \cite{kytola2009}. The problem of non-chiral logarithmic representation is even more complicated and remains widely open, see \cite{do2008,vasseur2012,ridout2012} for first attempts.

The previous scheme becomes degenerate for $r=1$, because there is no field $\xi_1$ below $\phi_1$ and $\psi_1$. The defining relations become ($h_{1,3}=0$)
\be
L_0 \psi_1 = \phi_1,  \qquad L_n \psi_r = 0 \;\;{\rm for\ }n \geq 1.
\ee
The graphical representation of ${\cal R}_{1,1}$ is still given in Figure \ref{Rr1} where the green region is to be removed. As a consequence, the parameter $\beta_1$ no longer makes sense.

An important issue concerns the fusions of these two sets of representations, namely the quotient ${\cal V}_{r,s}$ and the ${\cal R}_{r,1}$. It has been argued in \cite{gaberdiel1996} that the set of irreducibles ${\cal V}_{r,s}$ ($s \leq 2$) and the ${\cal R}_{r,1}$ {\it for specific values of $\beta_r$} is closed under fusion, with conjectured fusion rules given by
\bea
\V_{r_1,1} \star \V_{r_2,1} &=& \bigoplus_m \; \V_{m,1}, \qquad 
\V_{r_1,1} \star \V_{r_2,2} = \bigoplus_m \; \V_{m,2}, \label{VV} \\
\V_{r_1,2} \star \V_{r_2,2} &=& \bigoplus_m \; \R_{m,1}, \qquad 
\V_{r_1,1} \star \R_{r_2,1} = \bigoplus_m \; \R_{m,1},\label{VR} \\
\V_{r_1,2} \star \R_{r_2,1} &=& \bigoplus_m \; [\V_{m-1,2} \oplus 2\,\V_{m,2} \oplus \V_{m+1,2}], \\
\R_{r_1,1} \star \R_{r_2,1} &=& \bigoplus_m \; [\R_{m-1,1} \oplus 2\,\R_{m,1} \oplus \R_{m+1,1}],
\label{RR}
\eea
where the summations are  $|r_1-r_2|+1 \leq m \leq r_1+r_2-1$ by steps of 2, and $\V_{0,2}=\R_{0,1} \equiv 0$. One sees that all $\R_{r,1}$ appear in the fusion of irreducible representations and must therefore carry fixed values of $\beta_r$. The sequence of actual values starts with $\beta_r = -1,\, -18,\, -2700, \ldots$ for $r \geq 2$, where the general term is conjecturally given by $\beta_r = -(r-1) [(2r-3)!]^2 / 4^{r-2}$ \cite{vasseur2011}. These fusions have since been confirmed on the lattice in the dense polymer model, either in its loop formulation \cite{pearce2007} or in its (inequivalent) spin chain formulation \cite{read2007}. The fusions with the other, non-irreducible representations $\V_{r,s}$, apart from the $\V_{1,2r+1}$, see below, are not known.

Let us mention that the loop formulation of the dense polymer model gives rise to closely related representations, so-called Kac representations and noted $(r,s)$, for Kac labels $r,s \geq 1$ \cite{rasmussen2011}. For each pair $r,s$, the characters of $\V_{r,s}$ and $(r,s)$ are equal, but the two are nonetheless distinct, except in the following cases,
\bea
&& (r,1) = \V_{r,1}, \quad (1,2r) = (r,2) = \V_{1,2r} = \V_{r,2}, \quad{\rm (h.w.\ irreducible)} \\
&& (1,2r+1) = \V_{1,2r+1}. \qquad {\rm (h.w.\ reducible\ indecomposable)}
\eea
For $s$ even, $(r,s)$ is fully reducible and reduces to a direct sum of irreducibles, but in all other cases, $(r,s)$ is non highest weight, reducible and indecomposable of rank 1 ($L_0$ is diagonalizable). For the special value $s=3$, the representation $(r,3)$ for $r \geq 2$ is in fact isomorphic to the quotient $\R_{r,1}/\phi_r$ by the highest weight representation generated by $\phi_r$ (in the quotient, the diagonalizability of $L_0$ is recovered since $\psi_r$ and its descendants lose their partner, so that the Jordan cells all become rank 1).

All mixed fusions between Kac representations and the $\R_{r,1}$ for the special values of $\beta_r$ given above have been determined (conjectured) in \cite{rasmussen2011}. They allow to complete the fusions (\ref{VV})-(\ref{RR}) with
\bea
&& \V_{1,2r_1+1} \star \V_{1,2r_2+1} = \V_{1,2(r_1+r_2)+1} \oplus \bigoplus_{m=|r_1-r_2|+1,{\rm step\ }2}^{r_1+r_2-1} \; \R_{m,1},\\
&& \V_{1,2r_1+1} \star \V_{r_2,1} = (r_2,2r_1+1),\\
\noalign{\smallskip}
&& \V_{1,2r_1+1} \star \left\{\matrix{\V_{r_2,2} \cr \R_{r_2,1}}\right\}  = \bigoplus_{m=|r_2-r_1-{1 \over 2}|+{1 \over 2}}^{r_1+r_2} \; \left\{\matrix{\V_{m,2} \cr \R_{m,1}}\right\}.
\eea
Note the second line which closes on a single Kac representation. The set of representations $\{\V_{r,1}, \V_{r,2}, \V_{1,2r+1}, \R_{r,1}\}$ is therefore not closed under fusion.


\section{Lattice calculations}
\label{lattice}

From the general properties of the ASM reviewed in Section \ref{asm}, and in view of its relations to spanning trees and the loop-erased random walk, one may well be tempted to suspect that the sandpile model is critical, in two and higher dimensions, and therefore presumably conformally invariant. 

The criticality of the ASM appears to related to two essential ingredients, the nature of the recurrent configurations and the local conservation of sand in the model (except at a few dissipative sites). The recurrence property of the  configurations is shaped by the many topplings occurring in the system while approaching the stationary regime. Large scale topplings tend to organize far away sites in a specific way and therefore introduce large distance correlations. This mechanism however requires that large avalanches do occur, and this is only possible if most sites are conservative, enabling the excess of sand to be transported over large distances rather than being quickly evacuated to the sink through dissipative sites. It has indeed been shown that when all sites are dissipative (a uniform non-zero density is enough), the average avalanche size per added particle is finite and the correlations are exponential \cite{ghaffari1997,katori2000,mahieu2001,maes2004}. On the contrary, when the density of dissipative sites is zero, the average avalanche size diverges with the system size like $N^{2/d}$ in $d$ dimensions \cite{dhar1990}.

In this section, we will review the calculation of lattice correlations supporting the view that the ASM in two dimensions is critical and conformally invariant. Moreover we will collect enough evidence to argue that the underlying conformal description in the scaling limit is actually logarithmic. 

Before the correlations themselves, we will start by computing the marginal distributions of a few variables, mainly related to height variables. Although they are generally irrelevant for a field theoretic interpretation, it is instructive to see how they can be computed and in any case, they are useful to understand the calculation of the correlations themselves. 

Unless explicitly stated, all the calculations are made on $\Z^2$, mostly on rectangles and cylinders. We assume that the bulk sites are always conservative (except for a finite number of them when we consider the insertion of extra dissipation), whereas the boundary sites can be either conservative ($z_i^\star=z_i^{}=3$) or dissipative ($z_i^\star=z_i^{}+1=4$). In the former case, the site is called {\it closed}, and {\it open} in the latter; we will accordingly talk about closed and open boundary conditions. Then the toppling matrix is given by the Laplacian, plus appropriate boundary conditions, Neumann or Dirichlet for closed and open boundary conditions respectively. As we are primarily interested in the thermodynamic limit (followed by the scaling limit), all results will be given in the infinite volume limit. 

\subsection{Free energies}
\label{freeenergy}

Perhaps the most elementary result concerns the number of recurrent configurations, which provides a natural definition of the partition function,
\be
Z = \det \Delta.
\ee
At finite volume (i.e. on a finite graph), $Z$ is an integer. 

On a $L \times M$ rectangle, with the two opposite edges of length $L$ open and the other two closed, the partition function is given by
\be
Z_{\{{2\,{\rm open} \atop 2 \,{\rm closed}}\}} = \prod_{m=0}^{L-1} \prod_{n=1}^{M}
\, [4 - 2\cos{\textstyle {m\pi \over L}} - 2\cos{\textstyle {n\pi \over M+1}}].
\ee
In the limit $L,M \to \infty$ with $\tau = i{M \over L}$ fixed, the asymptotic behaviour of the free energy reads \cite{ruelle2002}
\be
\hspace{-2cm} 
\log{Z_{\{{2\,{\rm open} \atop 2 \,{\rm closed}}\}}} = L(M+1){4{\rm G} \over \pi} - (L + M + 1) \log(1+\sqrt{2}) + \half \log{M} + \log{2^{5 \over 4}\, \eta(\tau)} + \ldots
\label{freeen}
\ee
up to terms which vanish in the limit $L,M \to \infty$ ($\eta$ is the Dedekind eta-function). 

It follows from (\ref{freeen}) that the free energy per bulk site is equal to $4{\rm G} \over \pi$, and that the free energy per open resp. closed boundary site is ${6{\rm G} \over \pi} - {1\over 2}\log{(1+\sqrt{2})}$ and ${4{\rm G} \over \pi} - {1\over 2}\log{(1+\sqrt{2})}$. Exponentiating these numbers, we find, within the set of recurrent configurations, an effective number of degrees of freedom equal to 3.21, 3.70 and 2.07 for a bulk site, an open boundary site and a closed boundary site, to be compared with the values 4, 4 and 3 in the set of stable configurations.

The result (\ref{freeen}) also implies
\be
\lim_{m \to \infty} {1 \over M} \log Z = {4{\rm G} \over \pi}L - \log(1+\sqrt{2}) - {\pi \over 12L} + \ldots
\ee
The third term can be compared with the general result $\pi c \over 24L$ from conformal theory. It yields the first and simplest indication that the conformal description has $c=-2$ \cite{majdhar1992}.

\subsection{Local probabilities}
\label{local}

Beyond the free energy, a basic quantitative information is the distribution of the local values of the height variables, that is, the probability that a certain small cluster of fixed heights appears. On a finite lattice, these probabilities depend on the exact location of the cluster but become translation invariant in the thermodynamic limit. They retain however a dependence on whether the cluster is far or close to a boundary, and in the latter case, whether the boundary is open or closed. 

The simplest distribution is the probability $\P(h_i=a)$ that a given site $i$ has height $a$. 
Despite the apparent simplicity, these numbers are not so easily computed. To understand it, let us focus on the case when the reference site $i$ is in the bulk, far from the boundaries. The height can then take four values, from 1 to 4, so there are only four numbers to be computed (three really). The first question is: {\it how can we characterize the recurrent configurations that have a given height at a given site ?} 

The answer is rather impractical in terms of forbidden subconfigurations, and becomes much clearer in terms of spanning trees. To see this, we use the burning algorithm in a slightly different (but equivalent) way, by proceeding in two steps. We first start the algorithm as explained in Section 2, iteratively burning all burnable sites except the site $i$ at which we want to compute the height probability. When this is done, the recurrent configuration is only partially burnt, with a subset $\L_b$ of burnt sites and a complementary subset $\L_u$ of unburnt sites, which contains the site $i$ and possibly other sites. At this stage, $i$ must be burnable --it is even the only one in $\L_u$ to be so-- and will eventually propagate the fire to the whole of $\L_u$ (this is the second step). 

According to the burning algorithm, $i$ being burnable means that its height is strictly larger than the number of its unburnt nearest neighbours (those in $\L_u$). If this number is equal to $k$, between 0 and 3, we have $h_i > k$ and we say that $i$ has $k$ predecessors\footnote{A site $j$ is a predecessor of $i$ if the path going from $j$ to the root passes through $i$, so that $j$ is traversed first. By definition, the sites of $\L_u$ are the only predecessors of $i$.} among its nearest neighbours. Let $X_k(i)$ be the fraction of spanning trees such that the site $i$ has $k$ predecessors among its nearest neighbours.

The fraction $X_k(i)$ contributes equally to the values of $\P(h_i=a)$ for $k < a \leq 4$ because in a recurrent configuration with $h_i=4$ whose spanning tree is in $X_k(i)$, one can decrease $h_i$ from 4 down to $k+1$ and still keep it recurrent. Therefore one finds \cite{priezz1994}
\be
\P(h_i=a) = \sum_{k=0}^{a-1} \; {X_k(i) \over 4-k}. 
\label{singleh}
\ee

The number $X_0(i)$ is the easiest and can be computed in the following way \cite{majumdar1991}. The spanning trees counted by $X_0(i)$ have a single connection between $i$ and one of its four neighbours, which can be the N, W, S or E neighbour ($i$ is a leaf). By rotation invariance, which one it is does not matter in the thermodynamic limit, so we may focus on those trees in which $i$ is connected to its left neighbour say, and include a factor 4. The corresponding spanning trees can be viewed as {\it all} spanning trees on a modified lattice, obtained from the original one by removing the edges from $i$ to its N, W and S neighbours. In the modified lattice, the site $i$ has degree 1, while its three N, W and S neighbours have degree 3, so that the toppling matrix is accordingly modified by a rank-4 matrix $B$ as
\be
\textstyle \widetilde \Delta = \Delta + B[i], \qquad 
B[i] = {\scriptsize \pmatrix{-3 & 1 & 1 & 1 \cr 1 & -1 & 0 & 0 \cr 1 & 0 & -1 & 0 \cr 1 & 0 & 0 & -1}} \ {\rm on\ } \{i,{\rm N,W,S}\},
\label{blockB}
\ee
where $B$ is identically zero elsewhere. By Kirchhoff's theorem, $\det \widetilde \Delta$ gives the number of spanning trees in which $i$ is a leaf and is connected to its left neighbour, while the ratio $(\det \widetilde\Delta) / (\det\Delta)$ is their fraction in the full set of spanning trees. It follows from (\ref{singleh}) that
\be
\P(h_i=1) = {1 \over 4} X_0(i)= {\det \widetilde\Delta \over \det\Delta} = \det (\bi + \Delta^{-1}B[i]).
\label{det}
\ee

Because $B$ has rank 4, the previous determinant reduces to its four-dimensional restriction to \{$i$,N,W,S\}. In the infinite volume limit, $\Delta$ becomes the discrete Laplacian on $\Z^2$ while $\P(h_i=a) \to \P_a$ converges to a translation invariant quantity. The relevant entries of the Green matrix $\Delta^{-1}$ are known exactly, from which one obtains \cite{majumdar1991}
\be
\P_1 = {2(\pi - 2) \over \pi^3} \simeq 0.07363.
\label{P1}
\ee

The other three numbers $X_k(i)$ for $k \geq 1$ are not so easy. For spanning trees in $X_1(i)$ for example, only one neighbour of $i$ is a predecessor of $i$. So the fireline could come from the E neighbour of $i$, pass through $i$ and go down to its S neighbour, which would therefore be the predecessor of $i$. The two bonds from $i$ to its N and W neighbours can then not be part of the spanning trees in this class, and could be removed, like what we did for $X_0(i)$. However this is no longer sufficient because the fireline coming out from the S neighbour could wander in the lattice and eventually touch the W neighbour of $i$, in which case $i$ would have two predecessors at least.

We see that there is a clear, non-local distinction for a site to have no predecessor among its neighbours or to have $k \geq 1$ predecessors, because the fireline can go from $i$ to a neighbour directly, by using the edge connecting them, or along a path making a possibly long tour in the lattice. The spanning trees in $X_0(i)$ are constrained by local conditions, those in $X_{k \geq 1}(i)$ are subjected to non-local constraints. This simple observation makes the latter much more difficult to compute. In fact it took over twenty years before they could be given a simple closed form.

The first computation of $X_k(i)$ in the infinite volume limit was carried out by Priezzhev \cite{priezz1994} by purely graph-theoretic arguments. The result however was given in terms of two multiple integrals which could not be evaluated analytically. The computation was later reconsidered in \cite{jeng2006}, which established an exact linear relation between the two integrals. Based on a high-precision numerical evaluation of the remaining integral, the following expressions for 1-site probabilities were conjectured in \cite{jeng2006},
\bea
\P_2 &=& {1 \over 4} - {1 \over 2\pi} - {3 \over \pi^2} + {12 \over \pi^3} \simeq 0.1739,\\
\noalign{\smallskip}
\P_3 &=& {3 \over 8} + {1 \over \pi} - {12 \over \pi^3} \simeq 0.30629,\\
\noalign{\smallskip}
\P_4 &=& {3 \over 8} - {1 \over 2\pi} + {1 \over \pi^2} + {4 \over \pi^3} \simeq 0.44617.
\eea

Remarkably, the mean value of the height distribution at a site turns out to be
\be
\la h_i \ra = \sum_{a=1}^4 \: a\, \P_a = {25 \over 8},
\ee
a deceptively simple number in regard of the heavy technology used to obtain it, but definitely a hint that a simpler solution is lurking away.

Three independent proofs for the above three probabilities were eventually given within a year. The first one \cite{poghosyan2011/2} is based on a specific relation, noticed in \cite{poghosyan2011/1}, between the mean height $\la h_i \ra$ and the probability that a LERW eventually visits a fixed nearest neighbour of its starting point, the latter being then computed in terms of dimer arrangements and found to be $5/16$; the second one \cite{kenyon2011} also uses, in more general terms, the relation of height probabilities to LERW passage probabilities (see also \cite{levine2011} for a discussion of closely related quantities in spanning trees and spanning unicycles); finally the third one \cite{carac2012} relies on a clever and direct evaluation of the multiple integrals alluded to above. The reference \cite{kenyon2011} especially presents a much simpler scheme to compute the probabilities $\P_a$ (and more, see below), which provides an explanation for their simple form and the fact that they belong to $\Q[{1 \over \pi}]$. 

Beyond the height distribution at a single site, one can consider somewhat larger clusters, for instance the probability $\P(\twotwo)$ to have two adjacent heights 2. Generalizing the situation with the heights at a single site (1-site clusters), the larger clusters fall in two categories, those which can be computed in an elementary way, those which cannot.

The clusters which are {\it minimal subconfigurations} (aka {\it weakly allowed subconfigurations})  form the easy class: a given cluster of heights is called minimal if it becomes a forbidden subconfiguration (FSC) when one decreases the height of any site in the cluster \cite{majumdar1991,mahieu2001}. The smallest minimal subconfiguration is a single height 1, whose probability has been computed above, while larger examples include
\be
\hspace{-8mm} 
\conf1 \qquad\quad \conf2 \qquad\quad \conf3 \qquad\quad \conf6 \qquad\quad \conf9 \qquad\quad \conf{11}
\label{clusters}
\ee

The probability of occurrence of minimal subconfigurations can be computed in exactly the same way as the probability to find a height 1, by cutting off appropriate bonds and adjusting the degrees of the sites surrounding the cluster. They are thus all given by a determinant of the type (\ref{det}) where the form and the size of the matrix $B$ depends on the cluster considered. In the infinite volume limit, they are given by the determinant of a finite matrix involving entries of the Green matrix on $\Z^2$ (as such, they are in $\Q[{1 \over \pi}]$). For instance, the probability to find the fourth cluster in (\ref{clusters}) is given by \cite{mahieu2001}
\be
\hspace{-2.0cm}
\small 
-{1025 \over 128} + {1291 \over 8 \pi} - {10549 \over 8 \pi^2} + {151328 \over 27 \pi^3} - {353192 \over 27 \pi^4} + {429056 \over 27 \pi^5} - {634880 \over 81\pi^6} \simeq 0.000173.
\ee

The height clusters which are not minimal are in general much more difficult. Presently, the only practical method to compute the occurrence probabilities of these clusters --and their correlations on short distances-- is that of Kenyon and Wilson \cite{kenyon2011}, already mentioned above and based on the calculation of LERW passage probabilities (in addition the method works equally well on other lattices). Even so, the combinatorial complexity of the method increases very rapidly with the number of sites in the cluster, or with the distance separating the clusters in case of correlations. Using it, Wilson was able to compute the probability $\P(\abcd)$ to find the heights $a,b,c,d$ on four aligned adjacent sites and for any $a,b,c,d$, for instance \cite{wilson}
\bea
\hspace{-15mm}
\P(\2342) &=& {\textstyle {7042901 \over 256} - {322336899 \over 512\pi} + {28961498069 \over 4608\pi^2} - {30786770503 \over 864\pi^3} + {1294343142253 \over 10368\pi^4}} \nonumber\\
&& \hspace{-1.5cm} - {\textstyle {711072840871 \over 2592\pi^5} + {2817944526383 \over 7776\pi^6} - {2172951289219 \over 8748\pi^7} + {3564889622 \over 81\pi^8} + {52154353024 \over 2187\pi^9}}.
\eea

The variables we have discussed so far are related to heights. The description of the recurrent configurations in terms of rooted spanning trees offers other types of random variables. Let us recall that the oriented spanning trees can also be viewed as a loopless configurations of arrows, in which an outgoing arrow is attached to every site. With respect to the uniform distribution on all spanning trees, the probability that a certain local subconfiguration of arrows occurs can be computed.

To count the spanning trees with an arrow from $i$ to its neighbour $j$, we assign the bond $(i,j)$ a weight $\delta$ and take $\delta$ to infinity to give the other bonds incident to $i$ a relative weight equal to zero\footnote{We could assign a zero weight to the edges incident to $i$ other than $(i,j)$, but this is less economical.}. This can be implemented by defining a new (asymmetric) toppling matrix $\widetilde \Delta$, which coincides with $\Delta$ everywhere except for two entries: $\widetilde\Delta_{i,j} = -\delta$ (instead of $-1$) and $\widetilde\Delta_{i,i} = 3+\delta$ (instead of 4). By Kirchhoff's theorem, $\lim_{\delta \to \infty} {1 \over \delta} \det \widetilde\Delta$ is equal to the number of oriented spanning trees with an arrow going from $i$ to $j$. The new toppling matrix can be written as a finite rank perturbation of $\Delta$, as $\widetilde\Delta = \Delta + B[i,j]$ where the defect matrix is zero everywhere except on the sites $i,j$, where it reduces to the 2-block $\scriptsize \pmatrix{\delta-1 & -\delta+1 \cr 0 & 0}$. 

Likewise the determinant of the perturbed matrix $\Delta + B[i_1,j_1] + B[i_2,j_2] + \ldots$ computes the number of spanning trees containing arrows along the edges $(i_k,j_k)$. The probability to have $n$ arrows is then
\be
\P[{\rm arrows\ on\ } (i_k,j_k)] = \lim_{\delta \to \infty}\: {1 \over \delta^n} \, \det\Big(\bi + \Delta^{-1} \sum_{k=1}^n B[i_k,j_k]\Big).
\label{Parrows}
\ee
It reduces to the same sort of computations as for the minimal height subconfigurations, although fixing arrows between neighbouring sites does not univoquely fix the heights at those sites. For instance the probability, in the infinite volume limit, to have two right arrows on the edges $(i,i+\hat e_1)$ and $(i+\hat e_1,i+2 \hat e_1)$ is 
\be
\P(\circ \!\!\to \!\circ \!\!\to \!\circ) = {1 \over 4} - {1 \over 2\pi} \simeq 0.0908.
\ee
It shows that the configuration $\to\to$ is slightly more likely than $\to\uparrow$ or $\to\downarrow$, each of which has probability ${1 \over 4\pi} \simeq 0.0796$.

Finally let us mention that the probabilities discussed above are generally easier to compute on a boundary. In particular, the height distribution at a site on an open and on a closed boundary can be found in \cite{ivashkevich1994,piroux2005/1}.

\subsection{Dissipation}
\label{dissipation}

Closely related to the observables discussed in the previous section is the local {\it insertion of dissipation}. Because of the crucial r\^ole it plays in the understanding of the lattice correlations within a conformal picture, we give it a separate treatment. As we will see in a moment, it also yields the first explicit occurrence of logarithms. 

Let us consider a finite grid in $\Z^2$, with conservative bulk sites, the necessary dissipation being located on some part of the boundaries. We know that $\det \Delta$, with $\Delta$ the discrete Laplacian, counts the number of recurrent configuration and manifestly goes to infinity in the thermodynamic limit. We now ask the following question \cite{piroux2004}: what happens if we make a finite number of bulk sites dissipative ?

The dissipation rate at a site $i$ is controlled by the diagonal entry $\Delta_{i,i} = z_i^\star$ of the toppling matrix. A value $z_i^\star = z^{}_i$ ($=4$ for a bulk site) means $i$ is conservative, a larger value $z_i^\star > z^{}_i$ makes it dissipative. Thus a bulk site $i_1$ which is conservative in the original system becomes dissipative upon changing the diagonal entry by $\Delta_{i_1,i_1} \to  \Delta_{i_1,i_1}+1$, or $\Delta \to \Delta + D_{i_1}$ with $(D_{i_1})_{i,j} = \delta_{i,i_1}\delta_{j,i_1}$ (in fact we could shift $\Delta_{i_1,i_1}$ by any larger integer without any noticeable difference). Likewise the new toppling matrix $\widetilde \Delta = \Delta + D_{i_1} + D_{i_2} + \ldots$ modifies the original model by making the sites $i_k$ dissipative.

We expect that the insertion of dissipation at isolated sites, even in finite number, induces non-trivial changes over macroscopic scales. When the bulk is conservative, the spanning trees have an arrow flow oriented towards the boundary since the boundary sites are the only ones to be connected to the root. That some of the bulk sites become dissipative --connected to the root-- implies that the arrows can also flow to the root through them, which appear as sinks placed in the middle of the grid. The global pattern of arrows, or the structure of the spanning trees, is then dramatically different. 

A simple way to figure out the effect of introducing local dissipation is to see how the number of recurrent configurations changes, by computing the ratio $\det \widetilde\Delta / \det\Delta$, first at finite volume then in the thermodynamic limit.

Let us begin by introducing dissipation at a single site $i$, far from the boundaries. The perturbing matrix $D$ being of rank 1, the ratio of determinants reduces to
\be
{\det \widetilde\Delta_1 \over \det\Delta} = \det(\bi + \Delta^{-1}D_i) = 1 + \Delta^{-1}_{i,i}.
\label{n0}
\ee
The diagonal entry $\Delta^{-1}_{i,i}$ depends on the amount and location of dissipation in the unmodified model. It remains finite on a finite grid, but diverges in the infinite volume limit no matter what and where the dissipation is. This is to be expected because in the modified model, the height at site $i$ can take the extra value $h_i=5$, and this allows a large number of additional recurrent configurations on the grid. 

The same problem persists when we introduce dissipation at several sites. One possible solution is to compare the number of recurrent configurations with $n$ dissipative sites with the situation when there is one (rather than no) dissipative site. It is also natural to do so. The dissipation, if located only on the boundaries, becomes ineffective in the infinite volume limit because the boundaries are sent off to infinity. Thus one has to introduce dissipation by hand to make the model well-defined. Let us note that on the upper-half plane, the same remarks would apply if the boundary is closed, but not if the boundary is open.

Therefore we compare the partition functions with $n$ and with one dissipative site, and we compute $\det \widetilde\Delta_n / \det \widetilde\Delta_1$. These ratios make sense in the limit because the divergences of the numerator and the denominator are equal and proportional to $\Delta^{-1}_{0,0}$. Inserting dissipation at $n$ sites $i_1,i_2,\ldots$ leads to the ratios,
\be
{\det \widetilde\Delta_n \over \det \widetilde\Delta_1} = {\det\Big(\bi + \Delta^{-1} \sum_{k=1}^n D_{i_k}\Big) \over \det(\bi + \Delta^{-1} D_{i_1})} = {\det(\bi + \Delta^{-1})_{i,j \in \{i_1,\ldots,i_n\}} \over (1 + \Delta^{-1}_{i_1,i_1})}.
\label{deltan}
\ee
In the limit of infinite volume, this expression is actually symmetrical in the $i_k$'s since the denominator is translation invariant and does not depend on the position of $i_1$.

For $n=2$, the ratio reduces to an easy 2-by-2 determinant, and yields for a large distance $r = |i_1 - i_2|$ between the two insertions,
\be
{\det \widetilde\Delta_2 \over \det \widetilde\Delta_1} = {1 \over \pi} \log{r} + 2\gamma_0 + {\cal O}(r^{-2}),
\ee
where $\gamma_0 = {1 \over 2\pi}(\gamma + {3 \over 2} \log{2}) + 1$, and $\gamma = 0.577\, 216$ is the Euler constant.

These first results suggest a very neat conformal interpretation. Let us assume that in the scaling limit, the introduction of local dissipation at a bulk site $z$ is implemented by the insertion of a field $\omega(z,\bar z)$. The ratio (\ref{deltan}) would then converge to the $n$-point correlator $\la \omega(i_1) \ldots \omega(i_n) \ra$. Our prescription to normalize by $\det\widetilde\Delta_1$ and the result for $n=2$ imply
\bea
\la \omega(z,\bar z) \ra = 1,\\
\la \omega(z,\bar z)\,\omega(w,\bar w) \ra = {1 \over \pi} \log{|z-w|} + 2\gamma_0,
\eea
while the value for $n=0$, that is, no insertion at all, vanishes since it is the inverse of the diverging expression in (\ref{n0}), and leads to $\la \bi \ra = 0$.

The natural conclusion is that $\omega$ is a zero weight field, logarithmic partner of the identity $\bi$. The 2-point function fixes the conformal transformations of $\omega(z,\bar z)$ to be
\be
\hspace{-.5cm}
\omega(z,\bar z) \longrightarrow \omega(w,\bar w) - {1 \over 4\pi} \log{\Big|{{\rm d} w \over {\rm d} z}\Big|}^2\,, \quad {\rm or} \quad L_0\, \omega = \bar L_0\, \omega = - {1 \over 4\pi} \bi.
\label{transfomega}
\ee

Many more checks confirm this assignment \cite{piroux2004}. Similar calculations have been carried out on the upper-half plane for an open and a closed boundary, and in the latter case, dissipation on the closed boundary\footnote{An open boundary is dissipative in itself, so that adding extra dissipation does not change anything. Thus on an open boundary, the insertion of dissipation corresponds to the identity.} has been considered, leading to the same conclusion for a chiral boundary field $\omega_b(x)$. In addition various fusions have been considered: the bulk fusion of $\omega(z,\bar z)$ with itself, the corresponding boundary fusion of $\omega_b(x)$ and the way $\omega(z,\bar z)$ expands on $\omega_b(x)$ when it comes close to the closed boundary. Finally higher correlations with change of boundary conditions have also been computed. All of them are consistent with conformal correlators.

\subsection{Correlations in the bulk}
\label{correlations}

Long distance correlations are important to probe the critical properties of the model and to adjust its conformal description, but also raises new computational challenges. In this section, we review the known correlations of lattice variables located far from the boundaries, on the square lattice $\Z^2$. We mainly focus on the correlations of height variables, which are the most interesting, and briefly discuss the correlations of minimal subconfigurations and of arrow variables. We denote the general joint probabilities by $\P_{ab\ldots}(i_1,i_2,\ldots) \equiv \P[h(i_1)=a,h(i_2)=b,\ldots]$, but for the purpose of comparing with field-theoretic expressions, it is more convenient to use the subtracted random variables 
\be
h_a(i) \equiv \delta_{h(i),a} - \P_a(i), \qquad a=1,2,3,4,
\ee
and the corresponding subtracted correlations,
\be
\P^{\rm s}_{ab\ldots}(i_1,i_2,\ldots) = {\mathbb E}\left[h_a(i_1) \, h_b(i_2) \ldots\right].
\ee

We know that the height 1 is much easier to handle than the other three heights, so it is natural to start with the joint probabilities to have a height 1 at different sites. We have seen in a previous section that one can count the configurations with a height 1 at site $i$ by perturbing the toppling matrix by a 4-dimensional defect matrix $B[i]$, $\Delta \to \widetilde\Delta = \Delta + B[i]$, and computing its determinant. If we want to have a height 1 at different sites, we simply add as many blocks $B[i_k]$ as required. Therefore, the general height 1 joint probability is equal to the following finite determinant,
\be
\P_{11\ldots}(i_1,i_2,\ldots) = \det\Big(\bi + \Delta^{-1} \sum_k B[i_k]\Big).
\label{P11..}
\ee
The determinant is $4n$-dimensional for an $n$-site joint probability, but can be reduced to $3n$ by addition and subtraction of rows and columns. When the separation distances $\vec r_{kl}$ are large, one may expand the determinant in inverse powers of $r_{kl} = |\vec r_{kl}|$ by using the large distance expansion of the Green function on $\Z^2$ \cite{grigorev2009}. 

One finds that the 2-site correlation of two heights 1 decays algebraically 
\be
\P^{\rm s}_{11}(i_1,i_2) = -{\P_1^2 \over 2 r^4} - {4(\pi-2)[1 + (\pi-2)\cos{4\phi}] \over \pi^6 r^6} + \ldots
\label{P11}
\ee
where $\vec r_{12} = r \, {\rm e}^{\ci \phi}$. Only the first dominant term is rotationally invariant which suggests that the subtracted height 1 random variable $h_1(i)$ converges in the scaling limit to a scalar field with scaling dimension 2. In $d$ dimensions, the same 2-point correlation decays like $r^{-2d}$ \cite{majumdar1991}.

In complex notation $z_{kl}=\vec r_{kl}$, the 3-site correlation of three heights 1 is 
\bea
\hspace{-1cm} \P^{\rm s}_{111}(i_1,i_2,i_3) &=& {(\pi - 2)^2 \over \pi^9} \left\{{\pi-2 \over 2\,|z_{12}\,z_{13}|^4} + {2 \over z_{12}^3 \,z_{13}^3 \,\overline z_{23}^2} + {3(\pi-2) \over 2\,\overline z_{12}^2 \,z_{13}^2 \, z_{23}^4} \right. \nonumber\\
\noalign{\smallskip}
&& \hspace{3.5cm} \left. + {3(\pi-2) \over 2\,z_{12}^2 \,\overline z_{13}^2 \,z_{23}^4}+ {\rm perm.} + {\rm c.c.} \right\} + \ldots,
\eea
where the permutations to be added are the transpositions $(1 \leftrightarrow 2)$ and $(1 \leftrightarrow 3)$. The 4-site correlation is found to be given by, 
\bea
\hspace{-1.5cm} \P^{\rm s}_{1111}(i_1,i_2,i_3,i_4) &=& {\P_1^4 \over 8} \, \left\{{1 \over |z_{12}\,z_{34}|^4} + {1 \over |z_{13}\,z_{24}|^4} + {1 \over |z_{14}\,z_{23}|^4} - {1 \over (z_{12}\,z_{34}\,\overline z_{13} \,\overline z_{24})^2}\right. \nonumber\\ 
&& \left. - {1 \over (z_{13}\,z_{24}\,\overline z_{14} \,\overline z_{23})^2} - {1 \over (z_{14}\,z_{23}\,\overline z_{12} \,\overline z_{34})^2} 
+ {\rm c.c.}\right\} + \ldots
\eea
As we will see later, there is no need to compute higher correlations to get a convincing identification of the height 1 variable in the scaling limit.

Correlations involving heights not equal to 1 are much more complicated. To date, the only correlations that have been computed are $\P_{1a}(i_1,i_2)$ for $a=2,3,4$, and require heavy graph-theoretical and analytical calculations\footnote{This situation is very likely to change. Thanks to the technique developed in \cite{kenyon2011}, it is now possible to compute $\P_{1a}$ in a much more efficient way, on $\Z^2$ and on other lattices, and could even allow to compute higher correlations of the type $\P_{a1 \ldots 1}$ for $a>1$.}. Two-site correlations for two heights larger than 1 still remain well out of reach. We only quote here the final results for $\P_{1a}(i_1,i_2)$; the interested reader is referred to \cite{poghosyan2008,poghosyan2010} for the details.

\begin{figure}[t]
\hfill \includegraphics[width=90mm]{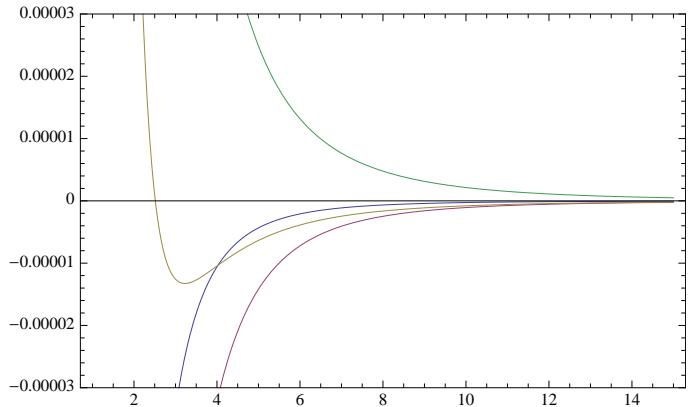}
\caption{\label{P1afig} Correlations $\P^{\rm s}_{1a}(r)$ as a function of $r$, for $a=1$ (blue), $a=2$ (purple), $a=3$ (gold) and $a=4$ (green).}
\end{figure}

The large distance behaviour of the subtracted lattice 2-site correlation is found to have the general form
\be
\P_{1a}^{\rm s}(r) = \frac{\alpha_a B + \beta_a A}{r^4} + \frac{\alpha_a A \log{r}}{r^4} + \ldots, \qquad 1 \leq a \leq 4,
\label{P1a}
\ee
up to lower order terms ${\cal O}(r^{-5}\log^k{r})$. The constants $A,B,\alpha_a,\beta_a$, which depend on the normalization of the lattice height variables and on the correlations themselves, can be computed analytically and take the following values
\bea
&& \textstyle A = -{\P_1^2 \over 2} = -{2(\pi-2)^2 \over \pi^6}, \quad B = -\frac{\P_1^2}{2} \left(\gamma + \frac{3}{2}\log{2}\right) -\frac{(\pi-2)(16-5\pi)}{\pi^6}, \\
\noalign{\smallskip}
&& \alpha_1 = 0, \quad \alpha_2 =  1, \quad {\textstyle \alpha_3 =  \frac{8-\pi}{2(\pi-2)}, \quad \alpha_4 = -\frac{\pi+4}{2(\pi-2)}}, \label{alphas} \\
\noalign{\smallskip}
&& \beta_1 = 1, \quad \beta_2 = 0, \quad {\textstyle \beta_3 = \frac{\pi^3 - 5\pi^2 + 12\pi - 48}{4(\pi-2)^2},\quad \beta_4 =  \frac{32 +  4\pi +  \pi^2 - \pi^3}{4(\pi-2)^2}}.
\label{betas}
\eea
Plugging these values into the general form (\ref{P1a}) yields the plots shown in Figure \ref{P1afig}. One sees in particular that the heights $a=1,2,3$ are anticorrelated with the height 1, implying that the heights 4 and 1 are positively correlated (since $\sum_a \P^{\rm s}_{1a}(r)= 0$). 

The previous results strongly suggest that the height 1 variable converges in the scaling limit to a primary field of dimension 2, while the other three height variables converge to a logarithmic partner of the same dimension. Additional calculations will confirm this picture.

Let us now briefly discuss the correlations of minimal subconfigurations, like those pictured in (\ref{clusters}). Their calculation follows closely that for heights 1. In particular the formula (\ref{P11..}) can again be used provided the various defect matrices are chosen according to the cluster variables one considers. In all cases, the correlation is expressed as a finite determinant, whose size increases with the number and the size of the clusters involved. These correlations have been extensively studied in \cite{mahieu2001} for 14 different clusters and their rotated versions. 

The correlation for any pair of minimal subconfigurations $S,S'$, for large separation, takes the general form
\be
\P^{\rm s}_{SS'}(z) = -{aa' \over 2\,|z|^4} - {b_1b_1' - b_2b_2' \over 4} \,\Big({1 \over z^4} + {1 \over \overline z^4}\Big) + \ldots
\label{SS'}
\ee
where the triplets of constants $(a,b_1,b_2)$ and $(a',b_1',b_2')$ are characteristic of the clusters $S,S'$. The dominant term given in the previous equation is not rotation invariant, which is not surprising in view of the fact that the clusters themselves are not all invariant. 

The correlations of minimal subconfigurations have also been computed in the dissipative perturbation of the ASM, with the advantage that they contain many more terms, allowing for a more reliable identification with fields in the scaling limit. A brief account of this is given below in Section \ref{mass}.

Finally the correlations between arrows follow the same calculational scheme as the minimal subconfigurations, as explained in Section \ref{local}, and are given by the similar expression (\ref{Parrows}). They are even easier to compute as the defect matrices are two-dimensional. The results are however different. For instance the (subtracted) correlations of two arrows, horizontal or vertical, separated by a large distance are given by the following expressions
\bea
&& P^{\rm s}_{\rightarrow\rightarrow}(z) = {1 \over 16\pi^2} \, \Big({1 \over z} + {1 \over \overline z}\Big)^2 + {\cal O}(z^{-4}),\\
\noalign{\smallskip}
&& P^{\rm s}_{\uparrow\uparrow}(z) = -{1 \over 16\pi^2} \, \Big({1 \over z} - {1 \over \overline z}\Big)^2 + {\cal O}(z^{-4}),\\
\noalign{\smallskip}
&& P^{\rm s}_{\rightarrow\uparrow}(z) = {\ci \over 16\pi^2} \, \Big({1 \over z^2} - {1 \over \overline z^2}\Big) + {\cal O}(z^{-3}).
\eea
The second correlation is obtained from the first one by a rotation by $\pi \over 2$ ($z \to -\ci z$).
Higher correlations of single arrows or correlations of larger local arrangements of arrows can be easily handled as well.

\subsection{Boundaries}
\label{bound}

Boundary conditions and their changes along a boundary, are important to see how a statistical system responds and is an efficient way to probe the content of its conformal description. We have so far mentioned two boundary conditions, open and closed. We start by examining them before introducing and discussing two more. We will restrict to the simplest geometric setting, namely the discrete upper-half plane $\Z \times \N$; we choose the boundary to be the real axis.

The difference between open and closed boundary conditions is in the toppling matrix: an open boundary contains open sites ($(\Delta_{\rm op})_{i,i} = z_i + 1 = 4$), a closed boundary contains closed sites ($(\Delta_{\rm cl})_{i,i} = z_i = 3$). As a consequence, $\Delta_{\rm op}$ is the Laplacian with Dirichlet boundary condition, $\Delta_{\rm cl}$ is the Laplacian with Neumann boundary conditions. 

It follows that one may ``close'' a number of sites in an open boundary by simply shifting the diagonal entries $(\Delta_{\rm op})_{i,i} \to (\Delta_{\rm op})_{i,i} - 1$ at those sites. The effect of closing a segment of an open boundary (and vice-versa) has been studied in \cite{ruelle2002}.

Thus we consider the ASM on the discrete upper-half plane with open boundary condition, characterized by a toppling matrix $\Delta_{\rm op}$. We close a segment $I$ of the boundary containing $n$ adjacent sites; the corresponding model with the new boundary condition is defined by the toppling matrix $\Delta_{\rm op}(n)$ given by
\be
\Delta_{\rm op}(n) = \Delta_{\rm op} - B, \qquad {\rm for\ } B_{i,j} = \delta_{i,j \in I}.
\ee
The defect matrix $B$ is the identity on the sites of $I$ and zero elsewhere.

The two partitions functions $Z_{\rm op}$ and $Z_{\rm op}(n)$, equal to the determinant of $\Delta_{\rm op}$ and $\Delta_{\rm op}(n)$ respectively, count the recurrent configurations for the two boundary conditions and are both infinite. The effect of closing the $n$ sites of $I$ can be measured by their ratio, which can be interpreted as the expectation value of the product of two operators, $\hat \phi^{\rm op,cl}$ and $\hat\phi^{\rm cl,op}$, flipping the boundary condition from open to closed and from closed back to open,
\be
{Z_{\rm op}(n\ {\rm cl}) \over Z_{\rm op}} = {\mathbb E}\left[\hat \phi^{\rm op,cl}(0) \, \hat\phi^{\rm cl,op}(n)\right] = {\det\Delta_{\rm op}(n) \over \det\Delta_{\rm op}} = \det(\bi - \Delta^{-1}_{\rm op}\,B).
\label{Zop}
\ee
This quantity, expected to be finite for finite $n$, should converge in the scaling limit to the field-theoretic 2-point function $\la \phi^{\rm op,cl}(0) \,\phi^{\rm cl,op}(n) \ra_{\rm CFT}$ of two boundary condition changing fields. Its exact value for $n$ large would therefore bring important information.

The obvious difficulty in computing the determinant (\ref{Zop}) is that its size grows with $n$. The specific form of the matrix $B$ however makes the calculation possible. As usual, the determinant reduces to the site indices at which $B$ is not zero, namely the sites in $I$ in the present case, where $B$ is simply the identity. Thus the determinant to be computed is $\det(\bi - \Delta^{-1}_{\rm op})_{i,j \in I}$. Because $\Delta_{\rm op}$ and its inverse are invariant under horizontal translations, the determinant has the Toeplitz form $\det (a_{k-\ell})$ for $k,\ell$ the horizontal coordinates of the sites in $I$. The enormous body of results on Toeplitz determinants can then be used.

One finds that the entries $a_m$ are the Fourier coefficients of the following periodic function,
\be
\sigma_1(k) = \sqrt{1-\cos{k}} \cdot \Big\{\sqrt{3-\cos{k}} - \sqrt{1-\cos{k}} \Big\}.
\ee
It has the form $\sigma_1(k) = (1-\cos{k})^\alpha \tau(k)$ where $\tau(k)$ is function that is single-valued, smooth, nowhere vanishing nor divergent on the unit circle. The asymptotic value of such a  determinant can be computed by using a generalization of the Szeg\"o theorem due to Widom \cite{widom1973}. The result reads \cite{ruelle2002}
\be
{Z_{\rm op}(n\ {\rm cl}) \over Z_{\rm op}} = \det(a_{k-\ell})_{1 \leq k,\ell \leq n} \simeq {\rm const.} \: n^{1/4} \, {\rm e}^{-{2{\rm G} \over \pi}n}, \qquad n\gg 1.
\ee
The exponential factor is related to the difference of boundary free energy between an open and a closed boundary site, equal to ${2{\rm G} \over \pi}$ as we have seen in Section \ref{freeenergy}. This non-universal factor must be omitted when comparing with the field-theoretic 2-point function, which becomes a pure power $n^{1/4}$. Thus the boundary condition changing field $\phi^{\rm op,cl} = \phi^{\rm cl,op}$ is expected to be a primary field a dimension $-{1 \over 8}$, degenerate at level 2. 

One may check this identification by considering the situation in which one closes two (well separated) segments on the open boundary. In this case the ratio of partition functions is related to the 4-point function of $\phi^{\rm op,cl}$, which can be determined exactly by assuming its degeneracy at level 2 (it is a complete elliptic function). On the lattice, the corresponding determinant is no longer Toeplitz and its asymptotic form cannot be computed analytically. The numerical evaluation of this determinant however compares very well with the conformal correlator (a complete elliptic function) and fully supports the above field identification \cite{ruelle2002}.

Let us also mention that the converse situation, namely we open segments in an otherwise closed boundary, can also be examined. Despite additional subtleties, it leads to identical results \cite{ruelle2002}. 

The open and boundary conditions are like free boundary conditions: they refer to the dissipative respectively conservative character of the boundary sites, but otherwise they allow the heights on the boundary to fluctuate freely (within the recurrent configurations). One could naturally think of some sort of fixed boundary conditions. In this class the simplest ones use again the language of spanning trees or arrows (remember that in the arrow picture, each site is assigned an outgoing arrow). The boundary conditions we want to consider force the arrows going out from boundary sites to be uniformly oriented, either to the left or to the right \cite{ruelle2007}. It turns out that this type of {\it windy boundary conditions} has a number of unusual features (like carrying an intrinsic orientation). It would too long to repeat the full analysis here, so we restrict to some illustrative steps.

To be concrete, let us start by inserting consecutive right arrows into a closed boundary. From Section \ref{local}, we know how to force a sequence of arrows. If we want a right arrow between two boundary neighbouring sites $i$ and $i+\hat e_1$, we simply add to the toppling matrix the 2-block $B[i,i+\hat e_1] = \scriptsize \pmatrix{\delta-1 & -\delta+1 \cr 0 & 0}$. If we impose $n$ consecutive right arrows between the sites $i, i+\hat e_1, i+2\hat e_1,\ldots ,i+n\hat e_1$, we form the matrix $B(\delta) = \sum_{k=0}^{n-1} B[i+k\hat e_1,i+(k+1)\hat e_1]$ and compute the following ratio, again interpreted as the expectation value of two operators which implement the two changes of boundary conditions,
\be
{Z_{\rm cl}(n\to) \over Z_{\rm cl}} = {\mathbb E}\left[\hat \phi^{\rm cl,\to}(0) \, \hat\phi^{\rm \to,cl}(n)\right] = \lim_{\delta \to \infty} \det\left(\bi + \Delta^{-1}_{\rm cl}\,B(\delta)\right).
\ee

It is not difficult to see that the limit over $\delta$ reduces to a finite $n \times n$ Toeplitz determinant, whose associated function,
\be
\sigma_2(k) = {\rm e}^{\ci (k-\pi)/2} \, \sqrt{1-\cos{k}} \cdot \Big\{\sqrt{3-\cos{k}} - \sqrt{1-\cos{k}} \Big\},
\ee
differs from $\sigma_1(k)$ used above by just a phase factor.

The phase makes a real difference though; the asymptotic behaviour of the associated determinant has been given in \cite{ehrhardt1997}. Using this result, one finds that
\be
{Z_{\rm cl}(n\to) \over Z_{\rm cl}} = {\rm const.\ } \, n^{-1/4} \, {\rm e}^{-{2{\rm G} \over \pi}n}, \qquad n\gg 1,
\label{Zcl}
\ee
where the exponential factor is again due to the difference of free energy between the two types of boundary conditions. This result suggests that the dimensions of the two fields $\phi^{\rm cl,\to}, \phi^{\rm \to,cl}$ add up to $-1/4$. If we restrict to the dimensions in the Kac table, the only possibility is that the two dimensions are $-1/8$ and $3/8$ (the two fields need not be equal as the arrows are outgoing from the closed boundary in one case, incoming in the other case). But with these two values, we seem to run into trouble since their conformal 2-point function vanishes identically. The apparent paradox can be solved if we remember the essential r\^ole of dissipation.

When explicit calculations are carried out on the infinite planar lattice, one generally starts with a finite system and takes the infinite volume limit of the results obtained at finite volume. As we know, this is well-defined provided the finite system involves dissipation, usually located on the boundaries. Even if the size of the system is ever increasing, dissipation is maintained at every step. However the conformal formulation on the full plane describes the system which is rightaway in its infinite volume limit, and so by itself is not aware that disspiation was present. Thus in order to include explicitly the presence of dissipation on the far away boundaries, one has to insert the field $\omega(\infty)$ representing the insertion of dissipation, as discussed in Section \ref{dissipation}.

In the present case, the boundary being fully closed, we must add dissipation by hand, so that the power law in (\ref{Zcl}) should be identified in the scaling limit with the 3-point function given by $\la \omega(\infty) \, \phi^{\rm cl,\to}(0) \, \phi^{\rm \to,cl}(n)\ra$. Its general form is indeed given by the correct power $\simeq n^{-1/4}$.

The same analysis in an open boundary leads to two further boundary condition changing fields, $\phi^{\rm op,\to}$ and $\phi^{\rm \to,op}$, both of dimension 0.

To go further, one may look at 3-point functions. Namely we change the boundary conditions three times by inserting two stretches of different conditions, of lengths say $n_1$ and $n_2$. The ratios of partition functions can be computed as above, by using the appropriate defect matrices, and all take the form of a finite determinants. As these do not have the Toeplitz form, they must be evaluated numerically. 

If we assume that the boundary conditions changing fields are primary, the generic form of a 3-point function is
\be
\la \phi_1(0) \, \phi_2(x) \, \phi_3(y) \ra = {\rm const} \, {(y-x)^{h_1-h_2-h_3} \over x^{h_1+h_2-h_3} \, y^{h_1+h_3-h_2}}.
\label{3pt}
\ee
By varying the lengths $x \equiv n_1$ and $y \equiv n_1+n_2$, a fit on the numerical data allows to determine the conformal dimension of each field. For this purpose, numerical calculations have been carried out for $n_1=20,30,50$ and $70$, and for each of these values, $n_2$ was varied from 10 to 150.

Two instructive situations are pictured below.

\begin{center}
\setlength{\unitlength}{0.8mm}
\begin{picture}(50,8)(3,-1)
\multiput(0,0)(2,0){8}{\line(1,0){1}}
\put(15,-1){\line(0,1){2}} \blue
\put(15,0){\vector(1,0){3}}
\put(18,0){\vector(1,0){3}}
\put(21,0){\vector(1,0){3}}
\put(24,0){\vector(1,0){3}}
\put(27,0){\vector(1,0){3}}
\put(30,0){\line(1,0){1}}\red
\put(31,0){\line(1,0){18}}\black
\put(31,-1){\line(0,1){2}}
\put(49,-1){\line(0,1){2}}
\multiput(49,0)(2,0){8}{\line(1,0){1}}
\put(7.5,-3){\makebox(0,0)[c]{\small op}}
\put(40.5,-2.5){\makebox(0,0)[c]{\small cl}}
\put(56.5,-3){\makebox(0,0)[c]{\small op}}
\put(23.5,2){\makebox(0,0)[b]{\small $n_1$}}
\put(40.5,2){\makebox(0,0)[b]{\small $n_2$}}
\end{picture}
\hspace{3cm}
\begin{picture}(50,8)(11,-1)
\multiput(0,0)(2,0){8}{\line(1,0){1}}
\put(15,-1){\line(0,1){2}}\blue
\put(15,0){\line(1,0){2}}
\put(19,0){\vector(-1,0){3}}
\put(22,0){\vector(-1,0){3}}
\put(25,0){\vector(-1,0){3}}
\put(28,0){\vector(-1,0){3}}
\put(31,0){\vector(-1,0){3}}\red
\put(31,0){\line(1,0){18}}\black
\put(31,-1){\line(0,1){2}}
\put(49,-1){\line(0,1){2}}
\multiput(49,0)(2,0){8}{\line(1,0){1}}
\put(7.5,-3){\makebox(0,0)[c]{\small op}}
\put(40.5,-2.5){\makebox(0,0)[c]{\small cl}}
\put(56.5,-3){\makebox(0,0)[c]{\small op}}
\put(23.5,2){\makebox(0,0)[b]{\small $n_1$}}
\put(40.5,2){\makebox(0,0)[b]{\small $n_2$}}
\end{picture}
\end{center}

For these two cases, the numerical data show that the related 3-point functions\footnote{Note that in the two cases, most of the boundary is open, i.e. dissipative, so there is no need to insert extra dissipation. Therefore in the conformal picture, the two situations are described by 3-point functions with no dissipation field $\omega$ inserted.} are consistent with the forms $({n_1+n_2 \over n_1 \sqrt{n_2}})^{1/2}$ and $n_2^{1/4}$ respectively. Comparing with (\ref{3pt}) determines or confirms the following weights,
\be
h^{\rm op,cl} = h^{\rm cl,\to} = -\textstyle{1 \over 8}, \qquad h^{\rm op,\to} = h^{\rm op,\leftarrow} = 0, \qquad h^{\rm cl,\leftarrow} = \textstyle{3 \over 8}. 
\ee

The last cases to consider is when the direction of arrows is changed, from right to left or vice-versa. The two changes of orientation are in fact different.

When right arrows are changed to left arrows, the two opposite arrows $\circ \!\!\!\to \!\circ \!\leftarrow \!\!\!\circ$ point to a same site $i$. Whether $i$ is open (connected to the root) or closed makes a substantial difference. Recalling that the arrows eventually flow towards the root, an open site $i$ allows the two sequences of arrows along the boundary pointing to $i$ to go directly to the root through $i$. If $i$ is closed, the flow has to go back in the bulk of the upper half-plane to find its way to the root. Thus the presence or absence of dissipation at the single site $i$ has strong effects on the global pattern of the arrow flow making up the tree. Thus we have to consider two different boundary condition changing operators, $\phi^{\to\stackrel{\rm op}{,}\leftarrow}$ and $\phi^{\to\stackrel{\rm cl}{,}\leftarrow}$.

When the arrows are changed from left to right $\!\circ \!\leftarrow \!\!\circ \: \circ \!\!\to \!\circ$, the previous duplication is not necessary since the sites concerned are not dissipative anyway (their arrow cannot point to the root by construction). Thus a single field $\phi^{\leftarrow,\to}$ is sufficient.

By following the same steps as above with 3-point functions, one obtains 
\be
h^{\to\stackrel{\rm op}{,}\leftarrow} = h^{\leftarrow,\to} = 0 ,\qquad h^{\to\stackrel{\rm cl}{,}\leftarrow} = 1,
\ee
completing the list of conformal weights for all possible fields switching among the four boundary conditions.

Beside their conformal weight, a much more important issue concerns the type of conformal representation these eight fields belong to. A convenient way to answer it is to use the constraints coming from the composition law of all these fields, expressed by the boundary fusion algebra. Namely, the fusion $\phi^{\alpha,\gamma} \star \phi^{\gamma,\beta}$ must close on fields which interpolate between the boundary conditions $\alpha$ and $\beta$, and should in particular contain the boundary condition changing field $\phi^{\alpha,\beta}$. 

Using these constraints, a proposal has been made in \cite{ruelle2007} that identifies the specific representations accomodating the eight fields, on the basis of two main assumptions: (i) the boundary condition changing fields are primary, degenerate at a level as low as possible, and (ii) they belong either to a highest weight representations ${\cal V}_{r,s}$ or to a rank 2 logarithmic representation $\R_{r,1}$ of the type reviewed in Section \ref{basics}, or to a quotient thereof. The results are summarized in Table 1.

\renewcommand{\arraystretch}{1.5}
\begin{table}[t]
\tabcolsep6pt
\hfill 
\begin{tabular}{|c||c|c|c|c|}
\hline
$\phi^{\alpha,\beta}$ & open & closed & $\to$ & $\leftarrow$ \\
\hline\hline
open & $\bi$ & \ $[-{1 \over 8}] \in {\cal V}_{1,2}$\ & $[0] \in {\cal V}_{1,3}$ & $[0] \in \R_{2,1}$\\
\hline
closed & \ $[-{1 \over 8}] \in {\cal V}_{1,2}$\ & $\bi$ &  $[-{1 \over 8}] \in {\cal V}_{1,2}$\ &\ $[{3 \over 8}] \in {\cal V}_{2,2}$\ \\
\hline
$\to$ &  $[0] \in \R_{2,1}$ & $[{3 \over 8}] \in {\cal V}_{2,2}$ & $\bi$ & 
$\renewcommand{\arraystretch}{1.2}
\begin{array}{c}
{}[0] \in \R_{2,1} \ ({\rm center\  op})\\
{}[1] \in \R_{3,1} \ ({\rm center\  cl})
\end{array}$ \\
\hline
$\leftarrow$ & $[0] \in {\cal V}_{1,3}$ & $[-{1 \over 8}] \in {\cal V}_{1,2}$ & $[0] \in {\cal V}_{1,3}$ & $\bi$ \\
\hline
\end{tabular}
\caption{Representations containing the fields $\phi^{\alpha,\beta}$ which implement a change of boundary condition from $\alpha$ (row label) to $\beta$ (column). The numbers in square brackets denote the scaling dimensions.}
\end{table}

To take an example, consider the field $\phi^{\rm op,\to}$. From the two assumptions, $\phi^{\rm op,cl}$ and $\phi^{\rm cl,\to}$ of weight $-1/8$ belong to an irreducible representation ${\cal V}_{1,2}$. Therefore $\phi^{\rm op,\to}$ should occur in the fusion $\V_{1,2} \star \V_{1,2} = \R_{1,1}$ (see (\ref{VR})). The representation $\R_{1,1}$ contains two fields of weight 0, a field $\phi_1$ that behaves like the identity, and its logarithmic partner $\psi_1$ (see Figure \ref{Rr1}). The field $\phi^{\rm op,\to}$ cannot be the identity, hence must be identified with $\psi_1$. However the primary partner $\phi_1$ must be null since the identity does not interpolate between two different boundary conditions. It follows that $\phi^{\rm op,\to}$ should belong to the quotient $\R_{1,1}/\phi_1 = {\cal V}_{1,3}$, and is degenerate at level 3 (it can be shown that its descendant at level 1 is not null \cite{ruelle2007}).

Table 1 can be completed using similar arguments \cite{ruelle2007}. The representations mentioned are those obtained by using the fusion rules as reviewed in Section \ref{basics}. Additional mixed higher correlators have been computed to cross-check the identifications of Table 1, namely 4-point correlators involving four changes of boundary conditions, or three changes of boundary conditions and the insertion of dissipation on a closed boundary, or of a boundary height 1 field. All of them have been found to be consistent with the proposal in Table 1.

However we cannot exclude the possibility that the fields in fact belong to quotients of these representations. Indeed we have observed in \cite{ruelle2007} that the primary fields $\phi_2$ and $\phi_3$ belonging respectively to $\R_{2,1}$ and $\R_{3,1}$ decouple in some correlators without being able to prove that they are actually null. If this turns out to be the case, the corresponding boundary condition changing fields would belong to quotients $\R_{2,1}/\phi_2$ and $\R_{3,1}/\phi_3$ which are no longer logarithmic, as explained in Section \ref{basics}.

On the other hand, if the fields $\phi_2$ and $\phi_3$ are not null, some of the boundary condition changing fields would belong to logarithmic representations, of rank larger than 1. The physical significance of this and the physical interpretation of the logarithmic fields remain open questions. 

\subsection{Boundary effects in height probabilities}

The height probabilities at a single bulk site have been discussed in Section \ref{local}. At an infinite distance from all boundaries, they are given by the four numbers $\P_a$ discussed in Section \ref{local}. Although they are probabilistically and combinatorially interesting, they are not so relevant for the conformal perspective. Much more important and relevant are the corresponding one-site probabilities at a finite but large distance to a boundary, especially if we impose different boundary conditions. 

In this section, we review the exact results for the 1-site height probabilities $\P_a(m)$ on the upper-half plane with open and closed boundary conditions, where $m$ is the distance to the boundary. Naturally the limit $\lim_{m \to \infty} \P_a(m)$ reproduces the numbers $\P_a$.

When the boundary condition is homogeneous on the real axis, these probabilities reflect the precise nature of the fields associated to the height variables since they correspond to 1-point functions of bulk fields, equivalently to 2-point functions of chiral fields. Moreover, the probabilities for open and closed boundary conditions are related to each other by the appropriate insertion of the boundary condition changing field $\phi^{\rm op,cl}$ and therefore provides a highly non-trivial check of the consistency of the conformal picture obtained so far for these fields. 

The graph theoretical techniques to compute the functions $\P_a(m)$ are basically identical to those used in the full plane, with two extra complications. First the full translation invariance is lost and reduces to the horizontal invariance only; this increases the number of graphs that need be handled separately. Second the functions $\P_a(m)$ are obtained as multiple integrals depending on the distance $m$ which require a rather long asymptotic analysis to compute the dominant contributions. 

The following results have been obtained in \cite{piroux2005/2,jeng2006} (and first in \cite{brankov1993} for $a=1$) when the boundary, here the real axis, is either open or closed,
\bea
\P_a^{\rm op}(m) &=& \P_a + {1 \over m^2} (c_a + {d_a \over 2} + d_a \, \log{m}) + \ldots,
\label{pop}\\
\P_a^{\rm cl}(m) &=& \P_a - {1 \over m^2} (c_a + d_a \, \log{m}) + \ldots,
\label{pcl}
\eea
up to terms of order ${\cal O}(m^{-3} \log^k{m})$. The coefficients are explicitly given by
\bea
c_1 &=& \textstyle {\P_1 \over 4} = {\pi - 2 \over 2\pi^3}, \qquad d_1 = 0, \label{a1b1}\\
\noalign{\smallskip}
c_2 &=& \textstyle {\pi - 2 \over 2\pi^3} \Big(\gamma + {5 \over 2} \log{2}\Big) - {11\pi - 34
\over 8\pi^3}, \qquad d_2 = {\pi - 2 \over 2\pi^3}, \label{a2b2} \\
\noalign{\smallskip}
c_3 &=& \textstyle {8 - \pi \over 4\pi^3} \Big(\gamma + {5 \over 2} \log{2}\Big) + {2\pi^2 + 5\pi
- 88 \over 16\pi^3}, \qquad d_3 = {8 - \pi \over 4\pi^3},\\
\noalign{\smallskip}
c_4 &=& -(c_1 + c_2 + c_3), \qquad d_4 = -(d_1 + d_2 + d_3).
\eea

One can make two observations regarding these results.

The first one concerns the form of these probabilities. The functions $\P_1(m)$, for both boundary conditions, are algebraic while all the others $\P_{a>1}(m)$ have an additional logarithmic term. It follows that $\P_3(m)$ and $\P_4(m)$ can be written, to order $m^{-2}$, as linear combinations of $\P_1(m)$ and $\P_2(m)$. Setting
\be
\P_a(m) - P_a = \alpha_a [\P_2(m) - P_2] + \beta_a [\P_1(m) - P_1],
\label{linear}
\ee
we obtain the coefficients 
\bea
&& \alpha_1 = 0, \quad \alpha_2 =  1, \quad {\textstyle \alpha_3 =  \frac{8-\pi}{2(\pi-2)}, \quad \alpha_4 = -\frac{\pi+4}{2(\pi-2)}},\\
\noalign{\smallskip}
&& \beta_1 = 1, \quad \beta_2 = 0, \quad {\textstyle \beta_3 = \frac{\pi^3 - 5\pi^2 + 12\pi - 48}{4(\pi-2)^2},\quad \beta_4 =  \frac{32 +  4\pi +  \pi^2 - \pi^3}{4(\pi-2)^2}},
\eea
that is, exactly those given earlier in (\ref{alphas}) and (\ref{betas}) ! We stress that the linear combinations (\ref{linear}) hold for either boundary condition {\it with the same coefficients}.

Second, the fact that the same set of coefficients $c_a,d_a$ control the probabilities for the two boundary conditions confirms that they are closely related. Precisely how depends on whether the open/closed condition changing field has been properly identified as a primary field of weight $-1/8$, but more importantly, on the exact nature of the fields which are supposed to describe the four height variables in the scaling limit. 

This should provide a crucial test to understand not only the above probabilities on the UHP, but also the bulk correlations reported in Section \ref{correlations} and the striking equality of the coefficients $\alpha_a,\beta_a$. We believe that the successful understanding of these specific feature constitutes one the most convincing support for the logarithmic conformal picture. The details for this are given in the next section.


\section{Conformal height variables}
\label{conform}

The natural microscopic random variables of the Abelian sandpile model are the height variables $h(i)$, assigned to the vertices and taking the four values $1,2,3,4$. The distribution of $h(i)$ at a single site and in the infinite volume limit can be computed exactly (see Section \ref{local}) but is of little value from the conformal point of view. Since conformal invariance enforces zero expectation values for the fields with non-zero scale dimension, one is more interested in the subtracted height variables. The joint probabilities we have computed so far show that the variables $h_a(i)$ pertaining to each of the possible height values at site $i$ make sense separately, and that they presumably converge in the continuum limit to conformal fields $h_a(z,\bar z)$,
\be
h_a(i) = \delta_{h(i),a} - \P_a \quad \longrightarrow \quad h_a(z,\bar z), \qquad a=1,2,3,4.
\ee
The main issue is now to guess enough of the conformal nature of the fields $h_a(z,\bar z)$ so as to reproduce all the lattice calculations reviewed in Section \ref{lattice}. 

From these, we learn a number of important facts: 
\begin{itemize}
\item all four fields have conformal weight (1,1), so a total scale dimension 2, 
\item $h_1$ is a non-logarithmic field, whereas the other three $h_{a>1}$ are logarithmic, 
\item looking back at (\ref{a1b1}) and (\ref{a2b2}), we observe the equality $d_2=c_1$, namely the logarithmic term of $\P_2(m)$ has the same coefficient as the algebraic term of $\P_1(m)$, a signal that $h_2(z,\bar z)$ could be the logarithmic partner of $h_1(z,\bar z)$, 
\item the linear relation (\ref{linear}), confirmed by the bulk correlations (\ref{P1a}), suggests that $h_3,h_4$ are linear combinations of $h_1,h_2$, namely $h_a(z,\bar z) = \alpha_a h_2(z,\bar z) + \beta_a h_1(z,\bar z)$.
\end{itemize}
It is therefore sufficient to find the conformal nature of two fields, $h_1(z,\bar z)$ and $h_2(z,\bar z)$.

The following conjecture has been first formulated in \cite{piroux2005/2}, and further analyzed in \cite{jeng2006}. It does not completely specify, and by far, the full representation to which $h_1,h_2$ belong, but rather sets the minimal framework needed to compute the relevant correlations. The conjecture is as follows:

\noindent
\vrule width 0.5mm depth 4.0cm
\vspace{-4.0cm}

\noindent
\leftskip=5mm
{\it The two bulk height fields $(h_1,h_2) = (\phi,\psi)$ form a logarithmic pair of weight (1,1). The field $\phi$ is primary and degenerate at level 2, while the conformal transformations of $\psi$ read
\bea
\hspace{-1cm} && L_0 \psi = \bar L_0 \psi = \psi + \lambda \phi, \quad L_1 \psi = \rho, \quad \bar L_1 \psi = \bar \rho, \quad L_{n>1} \psi = \bar L_{n>1} \psi = 0,
\label{conf1}\\
\noalign{\smallskip}
\hspace{-1cm} && L_{n \geq 0} \, \rho = \bar L_{n \geq 0} \, \bar \rho = 0, \quad \bar L_0 \rho = \rho, \quad L_0 \bar \rho = \bar\rho, \quad \bar L_{n \geq 1} \rho = L_{n \geq 1} \bar\rho = \kappa\, \bi \, \delta_{n,1},
\label{conf2}
\eea
where the two fields $\rho$ and $\bar\rho$ have weights (0,1) and (1,0) respectively, and $\kappa$ is a computable parameter determined by the normalization of $\psi$. }

\leftskip=0cm 
\bigskip
The previous transformations are all we need to compute the seeked correlations. They are reminiscent of the relations defining the representations $\R_{r,1}$ of Section \ref{basics}, and can be seen as a partial characterization of a non-chiral extension of $\R_{2,1}$ (the three fields $\psi,\phi,\rho$ satisfy the defining relations of $\R_{2,1}$ with respect to the left Virasoro modes, as do $\psi,\phi,\bar\rho$ with respect to the right modes). We will see that they are also consistent with the two additional relations
\be
L_{-1} \rho = \bar L_{-1} \bar \rho = \beta \lambda\, \phi.
\label{beta}
\ee
As in the chiral case, the parameter $\beta$ ought to be an intrinsic parameter that labels inequivalent representations \cite{ridout2012}. On the other hand, the parameter $\lambda$ is clearly related to the normalizations of $\phi$ and $\psi$, which are fixed if we insist that $\phi$ and $\psi$ are the scaling limit of the height 1 and height 2 lattice variables.

Altogether this yields seven fields $\{\bi,\rho,\bar\rho,\bar\partial\rho,\partial\bar\rho,\phi,\psi\}$ with total scale dimension less than or equal to 2. We stress again that this does not imply that these seven fields are the only ones on the lowest levels. 

On the basis of the assumptions given above about the conformal nature of the fields $\psi,\phi$, we can compute the correlators and compare with the results obtained on the lattice.

\subsection{Height variables on the upper-half plane}
\label{uhp}

The first correlators we would like to compute are $\la \mu^{\rm op,cl}(z_1) \, \mu^{\rm cl,op}(z_2) \, \phi(z,\bar z) \ra$ and $\la \mu^{\rm op,cl}(z_1) \, \mu^{\rm cl,op}(z_2) \, \psi(z,\bar z) \ra$, where $z_1,z_2 \in {\mathbb R}$ lie on the boundary whereas $z$ is in the bulk of the UHP. Since $\phi$ and $\psi$ are identified respectively with $h_1$ and $h_2$, the height 1 and height 2 variables, the two correlators, properly normalized by $\la \mu^{\rm op,cl}(z_1) \, \mu^{\rm cl,op}(z_2)\ra$, should correspond to the probability that the site at position $z$ has height 1 or 2 when the boundary condition on the real axis is all open except on the interval $[z_1,z_2]$ where it is closed. In particular the two limits 
\be
\hspace{-2.2cm} 
\lim_{z_1 \to z_2} {\la \mu^{\rm op,cl}(z_1) \, \mu^{\rm cl,op}(z_2) \, \psi(z,\bar z) \ra \over 
\la \mu^{\rm op,cl}(z_1) \, \mu^{\rm cl,op}(z_2)\ra} \quad {\rm and} \; \lim_{-z_1,z_2 \to \infty} {\la \mu^{\rm op,cl}(z_1) \, \mu^{\rm cl,op}(z_2) \, \psi(z,\bar z) \ra \over 
\la \mu^{\rm op,cl}(z_1) \, \mu^{\rm cl,op}(z_2)\ra}
\label{limits}
\ee
should reproduce respectively the functions $\P_2^{\rm op}(m)$ and $\P_2^{\rm cl}(m)$ given in (\ref{pop}) and (\ref{pcl}) for $z^{}-z^* = 2\ci m$. Similarly the same limits with $\psi$ replaced by $\phi$ ought to reproduce the functions $\P_1^{\rm op}(m)$ and $\P_1^{\rm cl}(m)$. Then $\P_a^{\rm op}(m)$ and $\P_a^{\rm cl}(m)$ for $a \geq 3$ will follow from the linear combinations $h_a = \alpha_a \psi + \beta_a \phi$.

Before going on, let us note that the two limits are in fact different. The first one leaves the boundary fully open, while it is fully closed in the second case. As follows from Section \ref{bound} where we have stressed the r\^ole of dissipation, the numerators should respectively correspond to $\la \psi(z,\bar z) \ra_{\rm op}$ and $\la \omega(\infty) \psi(z,\bar z) \ra_{\rm cl}$. This is not inconsistent with the fusion $\mu \star \mu \in \V_{1,2} \star \V_{1,2} = \R_{1,1}$ if we interpret it in two different ways. In the first limit, the boundary fusion $\mu^{\rm op,cl} \star \mu^{\rm cl,op}$ must close on fields living on an open boundary. We have noted before that the dissipation field on an open boundary corresponds to the identity, so that the fusion does not close on $\R_{1,1}$ but on the quotient $\R_{1,1}/\phi_1 = \V_{1,1}$. The second limit deals with the fusion $\mu^{\rm cl,op} \star \mu^{\rm op,cl}$ leaving a closed boundary on which the dissipation field is not trivial, so that this one really closes on fields in $\R_{1,1}$. Therefore we have \cite{piroux2004}
\bea
\mu^{\rm op,cl}(z) \, \mu^{\rm cl,op}(0) &=& z^{1/4} \, C_{\mu,\mu}^{\bi} \, \bi + \ldots \\
\noalign{\medskip}
\mu^{\rm cl,op}(z) \, \mu^{\rm op,cl}(0) &=& z^{1/4} \, C_{\mu,\mu}^\omega \,  [\omega(0) + \lambda \,
\bi \, \log{z}] + \ldots \qquad (\lambda=-{1 \over \pi}).
\eea
reinforcing the consistency of the conformal picture.

If the calculation of $\la \mu\mu\phi\ra$ is easy since both $\mu$ and $\phi$ are degenerate at level 2, that of $\la \mu\mu\psi\ra$ is tedious because the conformal transformations of $\psi$ are not homogeneous (both are non-chiral 3-point functions but reduce to chiral 4-point functions). It turns out that the most general form of $\la \mu\mu\psi\ra$ depends on 13 arbitrary coefficients, which can be reduced to 3 by imposing appropriate physical requirements\footnote{These are not related to the conformal nature of the field $\psi$: one requires that the limit $\lim_{z_1 \to z_2} z_{12}^{-1/4} \la \mu(z_1) \mu(z_2) \psi(z,z^*)\ra$ does not contain a logarithmic term $\log{z_{12}}$ and depends only on $z - z^*$, see \cite{jeng2006}.}. Two of them are related to the norm of $\psi$ and to the multiple of $\phi$ which can be freely added to it; when $\psi$ is identified as the scaling limit of the height 2 variable, these two coefficients are related to the numbers $c_2,d_2$ obtained from the lattice. Interestingly the third coefficient is related to $\beta$ through the relations (\ref{beta}).

For $\beta \neq 0$, one finds the following two correlators \cite{jeng2006}
\bea
&& \hspace{-2.0cm} \la \mu(z_1) \mu(z_2) \psi(z_3,z_4) \ra = {z_{12}^{1/4} \over z_{34}^2} \, {x-2
\over \sqrt{1-x}} \; \left\{2c_2 + 2d_2 \log\Big|{z_{34} \over 2}\Big| + {d_2 \over 2}
- {d_2 \over 4}  {x-2 \over \sqrt{1-x}} \right. \nonumber\\
\noalign{\medskip}
&& \hspace{5cm} \left. + \: d_2 {z_{34}^2 \over z_{13} z_{24}} \Big[{1 \over x-2} + {1 \over 2\sqrt{1-x}}\Big] \right\},\label{mumupsi}\nonumber\\
\noalign{\medskip}
&& \hspace{-1.5cm} + d_2 \, {1 - 2\beta \over 3\beta} \, {z_{12}^{1/4} \over z_{34}^2} \, {x-2
\over \sqrt{1-x}} \, \left\{1 + 3 \log{|z_{34}|} - 2 {\sqrt{1-x} \over x-2} - \log{(1 + \sqrt{1-x})^2 \over \sqrt{1-x}} \right\},
\eea
and
\be
\hspace{-2.0cm} \la \mu(z_1) \mu(z_2) \phi(z_3,z_4) \ra =  -{d_2 \over \lambda}\,{z_{12}^{1/4} \over 
z_{34}^2} {x-2 \over \sqrt{1-x}},
\label{mumuphi}
\ee 
where $z_{ij} = z_i - z_j$, $x = {z_{12} z_{34} \over z_{13} z_{24}}$ and $z_4 = z_3^*$. The constant $\kappa$ is equal to $\kappa = -d_2$.

Let us see how the above results reproduce the probabilities computed on the lattice. We set $z_3 = z$ and $z_4 = z^*$. 

We have to look at the limits when $z_1,z_2$ coincide, either at a finite point of the real axis, say the origin, or at infinity. So we set $-z_1 = z_2 = R$ and see what happens when $R \to 0$ or $+\infty$. From
\be
1 - x = {(R-z)(R+z^*) \over (R-z^*)(R+z)},
\ee
we see that $1-x$ has norm 1, and makes a full circle around 0 when $R$ takes the values from 0 to $+\infty$. This implies that $\sqrt{1-x}$ goes to $+1$ when $R \to 0$ and to $-1$ when $R \to +\infty$.
As the above correlators have this term as prefactor, one expects a change of sign in the two limiting cases, corresponding respectively to an open or a closed boundary. This is also the case in the probabilities given in (\ref{pop}) and (\ref{pcl}).

Another important consequence of the change of sign is that the very last term in (\ref{mumupsi}) is singular when $R\to \infty$, and brings a singularity $\log{z_{12}}$. As the limit of the correlator is supposed to reproduce $\P_2^{\rm cl}(m)$, which is perfectly regular, the singularity cannot be present. It either forces $d_2=0$, but then all logarithmic terms $\log{z_{34}} \sim \log m$ disappear, or else the value of the $\beta$ parameter is
\be
\beta_{\rm ASM} = {1 \over 2}.
\ee
With this value an easy calculation shows that the limits of $z_{12}^{-1/4} \la \mu(z_1) \mu(z_2) \psi(z,z^*)\ra$ in (\ref{limits}) exactly reproduce the two expressions (\ref{pop}) and (\ref{pcl}) for $a=2$.

The same limits of the other correlator yield
\be
\lim_{R \to 0 \; {\rm or} \; +\infty} z_{12}^{-1/4} \la \mu(-R) \mu(R) \phi(z,z^*)\ra = - {d_2 \over 2\lambda m^2} \;\;{\rm or} \;\; {d_2 \over 2\lambda m^2},
\ee
and again equal (\ref{pop}) and (\ref{pcl}) for $a=1$ provided $\lambda = -{1 \over 2}$.

To the best of our knowledge, the ASM provides the first lattice realization of a representation with the value $\beta={1 \over 2}$. It is well-known \cite{gaberdiel1999} that the bosonic sector of the symplectic free fermion theory contains fields $\rho_\theta,\bar\rho_\theta,\phi_\theta,\psi_\theta$ with identical conformal transformations to the above $\rho,\bar\rho,\phi,\psi$ but a different value of the parameter, namely $\beta_\theta=-1$. An extensive comparison between the two cases has been made in \cite{jeng2006}. The two values $\beta=-1,{1 \over 2}$ are precisely those which are discussed in the Example 7 of \cite{kytola2009}.

\subsection{Height variables across a strip}

The formula (\ref{mumupsi}) and (\ref{mumuphi}) allow to test the effect of a change of boundary condition and to compute finite-size corrections which can then be compared with the results of numerical simulations. 

A case which is particularly instructive is when the UHP is bordered by a boundary which is closed on the negative real axis and open on the positive part. Setting $z_1=-\infty$ and $z_2=0$ in (\ref{mumupsi}) and (\ref{mumuphi}) and taking $\beta={1 \over 2}$ and $\lambda=-{1 \over 2}$, we obtain the conformal prediction for the probability that the site $z$ in the bulk of the UHP has height $a$ in presence of a half-closed and half-open boundary,
\bea
\hspace{-1.5cm}
\P_a^{\rm cl|op}(z) - \P_a &=& \lim_{z_1 \to -\infty, \; z_2 \to 0} \; z_{12}^{-1/4} \: \la \mu^{\rm op,cl}(z_1) \, \mu^{\rm cl,op}(z_2) \, h_a(z,\bar z) \ra \nonumber\\
&=& -{z + z^* \over |z|(z-z^*)^2} \, \Big\{2c_a + 
2d_a \log\Big|{z-z^* \over 2}\Big| + {d_a \over 2} + {d_a \over 4} \,
{z + z^* \over |z|} \Big\},
\label{pmix}
\eea
where the coefficients $c_a,d_a$ are those appearing in (\ref{pop}) and (\ref{pcl}).

The conformal map $w = {L \over \pi} \log{z}$ transforms the UHP to an infinite strip of width $L$, under which the negative, closed, and the positive, open, parts of the real axis are mapped respectively onto the lines Im$\,w = L$ and Im$\,w=0$. The transformation of the (non-chiral) 3-point function $\la \mu \mu h_a \ra$ then allows to transport the previous probabilities to the strip. It requires the finite conformal transformation law of $\psi$, which can be computed by integrating the infinitesimal transformations (\ref{conf1}) and (\ref{conf2}). The result reads, for $\lambda = -{1 \over 2}$, 
\bea
\hspace{-2cm}
\psi_{\rm strip}(w,\bar w) &=& |z'(w)|^{2} \left\{\psi_{\rm uhp}(z,\bar z) 
- {1 \over 2} \log |z'(w)|^2 \, \phi_{\rm uhp}(z,\bar z) \right. \nonumber\\
&& \left. \hspace{1cm} + \; {z''(w) \over 2z'^{2}(w)}\, \rho_{\rm uhp}(z,\bar z)
+ {\bar z''(\bar w) \over 2\bar z'^{2}(\bar w)} \, 
\bar \rho_{\rm uhp}(z,\bar z) + \kappa \, \Big|{z''(w) \over 2z'^{2}(w)}\Big|^2\right\}.
\label{transfpsi}
\eea

Using this, we obtain the following expression for the probability on a strip with coordinate $w = u + \ci v$, $0 \leq v \leq L$,
\bea
\hspace{-1.5cm}
\P_a^{\rm strip}(w) - \P_a &=& \Big({\pi\over L}\Big)^{\!2} {\cos(\pi v/L)\over\sin^2(\pi
v/L)} \left\{ c_a + {d_a\over4} \Big[1+\cos\big({\pi v\over L}\big)\Big] \right.\nonumber \\
&& \hspace{4cm} \left. + \: d_a \log\!\Big[{L\over\pi}\sin\big({\pi v\over L}\big)\Big] \right\}
+ {d_a
\pi^2 \over 4L^2}.
\label{pistrip}
\eea
If the field identifications conjectured earlier are correct, this formula gives the (subtracted) probability, on an infinitely long strip of width $L$, that a site at position $w$ has a height value equal to $a$. By translation invariance, it only depends on the transverse coordinate $v$, which is also the distance of the site to the open edge, $L-v$ being the distance to the closed boundary.

The previous probabilities would not be easy to compute analytically on the lattice, but they can be estimated from numerical simulations (for a recent reviwe on numerical simulations in models like the ASM, see \cite{pruessner2013}). For these, the ASM dynamics is run on a finite rectangle of height $M$ and the four probability profiles are measured across the strip. For a ratio $M/L$ large enough, the numerical profiles should be good approximations of what they are on a infinite long strip, themselves presumably given by the above conformal predictions. To obtain the numerical data, about $10^{10}$ recurrent configurations have been sampled on a rectangle with $M=200$ and $L=50$ \cite{piroux2005/2}.

\begin{figure}[t]
\hspace{-0.5cm}\includegraphics[scale=0.63]{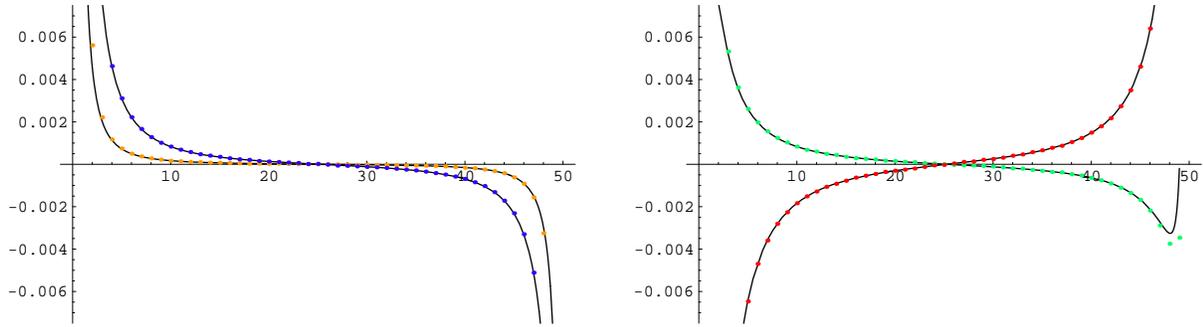}
\caption{These plots show the subtracted probabilities to find a height 1 (orange), 2 (in blue), 3 (green) or 4 (red) at a site lying along a transverse line of a rectangle $M \times L$, and
going from an open boundary (left end of the graph) to a closed boundary (right end).
The dots correspond to data obtained from numerical simulations, while the solid curves represent the conformal predictions (\ref{pistrip}). One notes the characteristic change of sign when going from an open to a closed boundary, discussed in Section \ref{uhp}.}
\label{Figure:simul}
\end{figure}

The results are given in the plots presented in Figure \ref{Figure:simul} and show a remarkable agreement. They provide very strong support to the conjecture identifying the four height variables $h_a(z,\bar z)$ as linear combinations of the two partner fields $\phi,\psi$, having the conformal transformations given in (\ref{conf1}), (\ref{conf2}).

\subsection{Height correlations on the plane}

Whether the field assignment as well as the conformal transformations conjectured for the height variables can reproduce the correlations computed on the plane is another consistency check. It is non-trivial for the following reason. From the discussion in Section \ref{bound} about the insertion of dissipation, it follows that the proper way to understand the lattice 2-height correlations computed on the plane is to say that they should be given, in the scaling limit, by a conformal 3-point function,
\be
\P_{ab}(z_{12}) - \P_a\P_b = \la h_a(z_1,\bar z_1) \, h_b(z_2,\bar z_2) \, \omega(\infty) \ra, \qquad |z_{12}| \gg 1.
\ee
Note the happy by-product that the correlation of two heights 1 is then not identically zero, as would have been the case had the field $\omega$ not been inserted.

When $a$ or $b$ is equal to 1, the probabilities $\P_{ab}(z_{12})$ are known to dominant order and given in (\ref{P1a}) so we can check whether their forms are consistent with the 3-point conformal correlators. For $a,b > 1$, the previous equation is a conformal prediction.

The 3-point correlator $\la h_a(z_1,\bar z_1) \, h_b(z_2,\bar z_2) \, \omega(z_3,\bar z_3) \ra$ can be computed from the conformal Ward identities or directly by applying the finite M\"obius transformation that maps the three points $z_1,z_2,z_3$ onto three fixed points, say $-1,+1,0$. From the transformations laws (\ref{transfomega}) and (\ref{transfpsi}) of $\omega$ and $\psi$, one easily obtains 
\bea
\hspace{-1cm} && \psi(z_1, \bar z_1) = {4 \over |z_{12}|^2}\Big\{\psi(-1) + \log{\Big|{z_{12} \over
2}\Big|}\, \phi(-1) - \rho(-1) - \bar\rho(-1) - \kappa \Big\},\\
\hspace{-1cm} && \psi(z_2, \bar z_2) = {4 \over |z_{12}|^2}\Big\{\psi(1) + \log{\Big|{z_{12} \over
2}\Big|}\, \phi(1) + \rho(1) + \bar\rho(1) + \kappa \Big\},\\
\hspace{-1cm} && \omega(z_3,\bar z_3) = \omega(0) - {1 \over 2\pi} \log{\Big|{z_{12} \over
2z_{13}z_{23}}\Big|}.
\eea
Inserting these in the correlator, one sees that it retains a logarithmic singularity when $z_3 \to \infty$ unless all 2-point functions of $\phi$ and $\psi$ vanish identically,
\be
\hspace{-1cm}
\la \psi(z_1, \bar z_1) \psi(z_2, \bar z_2) \ra = 
\la \phi(z_1, \bar z_1) \psi(z_2, \bar z_2) \ra = 
\la \phi(z_1, \bar z_1) \phi(z_2, \bar z_2) \ra = 0.
\ee
If the last equation is familiar, the first two are more peculiar and mark another difference with respect to fields $\phi_\theta$ and $\psi_\theta$ discussed earlier. 

One then obtains that the mixed correlators are related in the following way,
\bea
\hspace{-1cm} &&\la \phi(z_1,\bar z_1) \phi(z_2,\bar z_2) \omega(\infty) \ra = {A \over |z_{12}|^4}, \label{phiphi}\\
\hspace{-1cm} && \la \phi(z_1,\bar z_1) \psi(z_2,\bar z_2) \omega(\infty) \ra = 
{1 \over |z_{12}|^4} \Big\{B + A \log{|z_{12}|} \Big\},\\
\hspace{-1cm} && \la \psi(z_1,\bar z_1) \psi(z_2,\bar z_2) \omega(\infty) \ra = {1 \over |z_{12}|^4} \Big\{C + 2B \log{|z_{12}|} + A \log^2{|z_{12}|} \Big\}.
\eea
Taking the appropriate linear combinations yields the required correlators \cite{jeng2006},
\bea
&& \hspace{-1cm} \la h_a(z_1,\bar z_1) \, h_b(z_2,\bar z_2) \, \omega(\infty) \ra = {1 \over |z_{12}|^4} \Big\{\a_a\a_bC + (\a_a\b_b+\b_a\a_b)B + \b_a\b_b A \nonumber\\
&& \hspace{5mm}+\; [2\a_a\a_b B + (\a_a\b_b+\b_a\a_b)A] \log{|z_{12}|} +
\a_a\a_b A \log^2{|z_{12}|}\Big\}.
\label{pabgen}
\eea

For $b=1$ and from the values of the coefficients $\alpha_1=0, \beta_1=1$, it reduces to 
\be
\la h_a(z_1,\bar z_1) \, h_1(z_2,\bar z_2) \, \omega(\infty) \ra = {1 \over |z_{12}|^4} \Big\{ \a_a B + \b_a A + \a_a A  \log{|z_{12}|} \Big\}.
\ee
This exactly matches the form of the 2-site probabilities (\ref{P1a}) computed on the lattice. The three constants $A,B,C$ can only be obtained from lattice calculations; the coefficient $C$ is presently unknown.

Numerical simulations have been carried out to test the 2-site probabilities $\P_{ab}(r)$ when both $a,b$ are larger than 1. A fitted value of $C \sim -0.009$ shows an excellent agreement, even on surprisingly small distances (see  \cite{jeng2006} for plots). 

Let us recall from Section \ref{correlations} that when they involve the height 1 variable only, more correlators are known. In fact it has been observed \cite{mahieu2001} that the scaling limit of the 2-, 3- and 4-site probabilities are all correctly reproduced if the height 1 variable is identified in the scaling limit with the following composite field in the symplectic free fermion theory \cite{gurarie1993,kausch2000},
\be
h_1(z,\bar z) = \phi_\theta(z,\bar z) \equiv -\P_1 :\!\partial\theta \bar\partial\tilde \theta + \bar\partial\theta \partial\tilde \theta\!:\,, \qquad \P_1 = {2(\pi - 2) \over \pi^3},
\label{h1symp}
\ee
provided the zero weight field $\omega_\theta = :\!\theta\tilde \theta\!\!:$, which plays the r\^ole of dissipation, is inserted in the field theoretic correlator. In particular the 4-point correlation $\la \phi_\theta(1) \phi_\theta(2) \phi_\theta(3) \omega_\theta \ra = 0$ vanishes identically, in agreement with the result that the lattice 3-site joint probability decays with a global power 8 in the distances (rather than a power 6), implying that its scaling limit vanishes.

More generally, it has been shown \cite{mahieu2001} that the (subtracted) variables related to minimal subconfigurations $S$ can all be identified with the following combination of fields of weights (1,1), (2,0) and (0,2), for appropriate coefficients $a,b_1,b_2$ depending on $S$,
\be
\hspace{-1.5cm} 
h_S(z, \bar z) = -\left\{a :\!\partial\theta \bar\partial\tilde \theta + \bar\partial\theta \partial\tilde \theta\!: + \: b_1 :\!\partial\theta \partial\tilde \theta + \bar\partial\theta \bar \partial\tilde \theta\!: + \: \ci b_2 :\!\partial\theta \partial\tilde \theta - \bar\partial\theta \bar \partial\tilde \theta\!: \right\}.
\label{generaltheta}
\ee
For each of about a dozen minimal subconfigurations, the three coefficients have been determined exactly, and all mixed 2-point correlators have been computed. They all match the lattice result given in (\ref{SS'}). This conclusion holds in much greater generality, since it has been proved in \cite{jeng2005/1} that the variable associated to any local bond modification converges in the scaling limit to a field of the form (\ref{generaltheta}). A more direct derivation of these fields from the defect matrix used to compute the correlations has been given in \cite{moghimi2005}.

Thus even though the variables associated with the minimal subconfigurations, including the height 1 variable, can be consistently described within the free symplectic fermion theory, it turns out that this is no longer the case for the higher height variables which force a different value for the parameter $\beta$.


\section{Other developments}

In this last Section, we would like to briefly mention a number of complementary results and further developments. In most cases they complete the current picture we have of the Abelian sandpile model.

\subsection{The massive sandpile model}
\label{mass}

In the lattice calculations we have presented so far, the dissipation is localized on the boundaries or at isolated sites in the bulk. In all these cases, the density of dissipative sites vanishes in the infinite volume. We have argued that this makes the system correlated over large distances, with the consequences that the avalanche size distribution after the addition of a single grain has a power-lawed tail, and the correlations decay algebraically.

As mentioned before, the situation is drastically different when all sites, or at least a non-zero density, are dissipative. It is known in this case that the average size of the avalanches is finite and that the correlations decay exponentially \cite{ghaffari1997, katori2000, mahieu2001, maes2004}. 

The dissipative model is defined simply by changing the diagonal entries of the toppling matrix. The simplest possibility is to choose $\Delta_{i,i} = z^\star_i = z^{}_i+t$ for all sites, including the bulk sites. As compared to conservative bulk sites, for which $z^\star_i = z^{}_i$, the height variables now take $t$ extra values, namely they range between 1 and $z^{}_i+t$ (so between 1 and $4+t$ on a square grid), and more importantly, each time a site topples, $t$ grains of sand exit the system. This causes a substantial loss of sand during the relaxation process, and thus weakens the avalanches which then travel over typically much smaller distances. The conservative model is recovered in the limit $t \to 0$.

The model with $t>0$ is non-critical. The correlations are controlled by the inverse of the toppling matrix, a discrete massive Laplacian, whose large distance behaviour is given by the modified Bessel function $K_0(r\sqrt{t}) \sim e^{-r\sqrt{t}}$, implying that the correlation length diverges at $t=0$ like $\xi \sim {1 / \sqrt{t}}$. The scaling regime is reached by taking simultaneously the large distance limit $r={z \over a} \to \infty$ and the critical limit $t=a^2M^2 \to 0$ in terms of a scale $a \to 0$. The product $r\sqrt{t} \to Mz$ then defines an effective mass and a macroscopic distance. Thus the insertion of dissipation everywhere in the sandpile model is like a thermal perturbation away from the critical point, and corresponds to a massive theory. Around the critical point, the average avalanche size is finite and behaves like $1/t$ \cite{katori2000}.

The correlations of minimal subconfigurations in the dissipative model have been computed on the square lattice \cite{mahieu2001}. For instance, the correlator of two heights 1 is given at dominant order by
\be
\hspace{-1cm}
\P_{11}^{\rm s}(r) = -t^2\,\P_1^2 \left\{ {1 \over 2} K_0''^2 - {1 \over 2} K_0\,K_0'' + 
{1 \over 2\pi} K_0'^2 + {1+\pi^2 \over 4\pi^2} K_0^2\right\} + \ldots
\label{2ptmass}
\ee
up to higher order terms in $t$, and where the argument of all Bessel functions is $r\sqrt{t}$. In the prefactor, $\P_1$ is the height 1 probability at the critical point.

As the height 1 has the field identification (\ref{h1symp}) in terms of massless symplectic fermions, it is tempting to extend it to the massive case. Indeed the scaling limit of the correlator (\ref{2ptmass}) exactly matches the 2-point function of 
\be
\phi_\theta(z,\bar z) = -\P_1 \left[ :\!\partial\theta \bar\partial\tilde \theta + \bar\partial\theta \partial\tilde \theta\!: \,+\, {M^2 \over 2\pi}:\!\theta\tilde \theta\!: \right], 
\ee
where the two symplectic fermions are now massive and described by the action $S = {1 \over \pi} \int \partial\theta \bar\partial\tilde \theta + {M^2 \over 4} \theta\tilde \theta$, with 2-point function $\la \theta(z)\tilde \theta(w) \ra = K_0(M|z-w|)$.

Similar results have been obtained for the variables related to minimal subconfigurations \cite{mahieu2001}; nothing is known however of the massive extension of the higher height variables $h_2,h_3$ and $h_4$.

\subsection{Boundary height variables}

We have discussed at length the height variables on the plane or on the upper half-plane, to find that in the bulk, the height 1 has a very different scaling behaviour compared to the height 2, 3 and 4 variables. A natural question concerns the nature of the height variables on a boundary. This has examined for open and closed boundaries, in \cite{ivashkevich1994,jeng2005/2} for the critical model and in \cite{piroux2005/1} for the massive model.

That fact that the height correlations can be computed much more easily on a boundary than in the bulk has been noticed in \cite{ivashkevich1994}, where 2-site correlators have been first obtained in the critical model. However they do not provide enough information to decide whether or not the different height variables have different scaling properties. This question was settled almost simultaneously in \cite{jeng2005/2} by the calculation of multisite correlators in the critical case, and in \cite{piroux2005/1} by looking at 2-site correlators in the massive model, which allows for much finer field identifications.

The results depend on the boundary condition, open or closed, since the height variables take a different number of values, given in the critical case by $4$ and $3$ respectively. The main conclusion is that, for both boundary conditions, none of the height variables is logarithmic. Again the field identifications can be more conveniently expressed within the symplectic free fermion theory. We simply quote the results in the critical model.

For an open boundary, one finds that all four height variables are proportional to the same field,
\be
h^{\rm op}_a(x) = N^{\rm op}_a \, :\!\partial\theta \partial\tilde \theta \!:\,, \qquad a=1,2,3,4,
\ee
where the normalizations $N_a^{\rm op}$ are known exactly, and where the fields $\theta,\tilde \theta$ satisfy Dirichlet boundary conditions. 

On a closed boundary, the three height variables converge to different scaling fields,
\be
h^{\rm cl}_a(x) = N^{\rm cl}_a \, :\!\partial\theta \partial\tilde \theta \!: \,+\, M^{\rm cl}_a \,:\!\theta \partial\partial\tilde \theta \!: \,, \qquad a=1,2,3,
\ee
for known coefficients $N^{\rm cl}_a, M^{\rm cl}_a$, and where the fermions satisfy Neumann boundary conditions. In particular $M^{\rm cl}_1=0$ so that the height 1 variable behaves the same way on the two types of boundaries.

In the critical, non-dissipative model, all mixed 2-point correlators of boundary height variables decay like $r^{-4}$.

\subsection{Other lattices}

The sandpile can be defined and studied on any type of lattice, but the most interesting examples are arguably the two-dimensional regular lattices. Among these, the triangular or the honeycomb lattices are natural to test the universality properties of the model. Little has been done so far.

Numerical simulations have been carried out on the square, honeycomb and triangular lattices to determine the spectrum of exponents of toppling waves, which showed that the exponents are identical for the three types of lattices \cite{hu2003}.

Correlations of heights 1 have been computed exactly on the honeycomb lattice in \cite{azimi2010}. The results confirm that the height 1 scales in exactly the same way as on the square lattice, in the bulk and on an open or closed boundary. Only the lattice normalizations differ. 

The techniques developped recently in \cite{kenyon2011} to compute LERW passage probabilities offer a very interesting perspective since they can also be used on the triangular and honeycomb lattices. The results contained in \cite{kenyon2011} already yield the height distribution at one site. On the honeycomb lattice for instance, they read (each site in the bulk has three neighbours so that the heights take three values)
\be
\P_1 = {1 \over 12}, \qquad \P_2 = {7 \over 24}, \qquad \P_3 = {5 \over 8}.
\ee

The calculation of 2-point correlations involving a height 1 and a higher height should be possible using these techniques and would presumably confirm the type of scaling found on the square lattice, namely that height 1 is not logarithmic while the higher heights are. These computations remain to be done.

\subsection{Conformal invariance from SLE}

In recent years, the search of conformally invariant properties has been generalized to extended geometrical objects, like certain clusters and interfaces. In this respect the most successful approach is the one based on the Stochastic Loewner Equation (SLE), which has led to an enormous body of results, see for instance the reviews \cite{kager2004,cardy2005}.

These ideas have been applied to the ASM on the square lattice in \cite{saberi2009} by looking at the avalanche clusters. More precisely, after the addition of a grain, an avalanche may take place, during  which a certain number of sites topple, usually several times. The set of sites which topple at least once form the avalanche cluster. By generating these clusters numerically, the geometric features of the avalanche cluster boundaries can be studied. 

It has been argued \cite{saberi2009} that the avalanche frontiers are random curves with a fractal dimension numerically very close to 5/4, the same value as the loop erased random walks. In addition the statistics of these curves appears to be consistent with a description by an SLE process for the value of the parameter $\kappa=2$, corresponding to a value $c=-2$ of the central charge. The traces of SLE$_{\kappa=2}$ are known to be statistically identical to the LERW \cite{lawler2004}.

Similar simulations on the honeycomb lattice \cite{azimi2010} lead to the same conclusion, confirming its universality.

It is important to note that the clusters examined here belong to the dynamical aspects of the ASM, since they appear during the relaxation process. Consequently they are not the kind of random variables whose statistics are in principle determined by the stationary measure.


\section{Conclusion and perspectives}

In this review, we have collected the most important results obtained in the stationary regime of the Abelian sandpile model, in connection with the issue of conformal invariance. The main results concern: the height variables, in the bulk and on boundaries, different types of boundary conditions and the fields effecting a change of boundary conditions, the crucial r\^ole of dissipation. A large number of mixed correlation functions have been calculated on the lattice, either analytically or numerically, and perfectly match the conformal predictions.

The fairly safe conclusion we may draw from these data is that the ASM in its stationary regime is a conformal invariant system, and that it provides a lattice realization of a logarithmic conformal theory. We believe that it is one of the few lattice models where the logarithmic conformal invariance, with its strange and unusual features, can be best understood, and is a rare example where correlation functions can be explicitly computed.

Our current understanding of the model from the conformal point of view remains nonetheless very lacunar. The main question of course is to improve our knowledge of the spectrum of representations present in the conformal theory, both bulk and boundary, and to continue to investigate their precise nature and their fusions. The task can be tough; we have seen no sign so far of an extended symmetry, meaning that the relevant conformal theory could very well be irrational. Even though certain observables are correctly accounted for within the free symplectic fermion theory, the conformal nature of the bulk height variables suggests that the latter is in fact not the correct theory to describe the physics of the ASM.


\ack

As Senior Research Associate, I thank the Belgian National Fund for Scientific Research (FNRS) and further acknowledge the support of the Belgian Interuniversity Attraction Poles Program P7/18 through the network DYGEST (Dynamical, Geometry and Statistical Physics). Over the years, I have greatly benefited from discussions and collaborations with Deepak Dhar, Michael Flohr, Matthias Gaberdiel, Jesper Jacobsen, Kalle Kyt\"ol\"a, Paul Pearce, Vahagn Poghosyan, Vyatcheslav Priezzhev, Jorgen Rasmussen, David Ridout, Hubert Saleur and David Wilson.


\end{document}